\documentclass[apj]{emulateapj}
\pdfoutput=1
\usepackage{amsmath}
\usepackage{mathrsfs}
\usepackage{lscape}
\usepackage{longtable}
\renewcommand{\vec}[1]{\mbox{\boldmath$#1$}}
\newcommand{\thco}{$^{13}$CO}
\newcommand{\ceo}{C$^{18}$O}
\newcommand{\nht}{NH$_3$}
\newcommand{\hii}{H\,{\sc ii}}
\newcommand{\kms}{km\,s$^{-1}$}
\newcommand{\msun}{$M_{\odot}$}
\newcommand{\lsun}{$L_{\odot}$}
\newcommand{\eg}{{\it e.g.}}
\newcommand{\core}{L1551-MC}
\newcommand{\cmt}{cm$^{-3}$}
\newcommand{\uv}{$u$--$v$}
\newcommand{\irsf}{L1551~IRS5}
\newcommand{\lne}{L1551~NE}
\newcommand{\hr}{Hertzsprung-Russell}
\newcommand{\hone}{H\,{\sc i}}
\begin{document}
\slugcomment{Accepted to the Astrophysical Journal Supplemental Series}
\shorttitle{A Case Study of Low-Mass Star Formation}
\shortauthors{Swift \& Welch}
\title{A Case Study of Low-Mass Star Formation}
\author{Jonathan J. Swift}
\affil{Institute for Astronomy, 2680 Woodlawn Dr., Honolulu, HI
96822-1897: \tt{js@ifa.hawaii.edu} }

\author{William J. Welch}
\affil{Department of Astronomy and Radio Astronomy Laboratory,
University of California, 601 Campbell Hall, Berkeley, CA
94720-3411}

\begin{abstract}
This article synthesizes observational data from an extensive program
aimed toward a comprehensive understanding of star formation in a
low-mass star-forming molecular cloud. New observations and published
data spanning from the centimeter wave band to the near infrared
reveal the high and low density molecular gas, dust, and pre-main
sequence stars in L1551.

The total cloud mass of $\sim 160$\,\msun\ contained within a 0.9\,pc
has a dynamical timescale, $t_{\rm dyn} = 1.1$\,Myr. Thirty-five
pre-main sequence stars with masses from $\sim 0.1$ to 1.5\,\msun\ are
selected to be members of the L1551 association constituting a total
of $22\pm5$\,\msun\ of stellar mass. The observed star formation
efficiency, ${\rm SFE} = 12$\%, while the total efficiency,
${\rm SFE}_{\rm tot}$, is estimated to fall between 9 and 15\%. 

L1551 appears to have been forming stars
for several $t_{\rm dyn}$ with the rate of star formation increasing
with time. Star formation has likely progressed from 
east to west, and there is clear evidence that another star or stellar
system will form in the high column density region to the northwest of
\irsf.

High-resolution, wide-field maps of L1551 in CO isotopologue emission
display the structure of the molecular cloud at 1600\,AU physical
resolution. The \thco\ emission clearly reveals the disruption of the
ambient cloud by outflows in the line core and traces the interface
between regions of outflow and quiescent gas in the line
wings. Kinetic energy from outflows is being deposited back into the
cloud on a physical scale $\lambda_{\rm peak} \approx 0.05$\,pc at a
rate, $\dot{E}_{\rm input} \approx 0.05$\,\lsun. The remaining energy
afforded by the full mechanical luminosity of outflow in L1551
destroys the cloud or is otherwise lost to the greater interstellar
medium. 

The \ceo\ emission is optically thin and traces well the
turbulent velocity structure of the cloud. The total turbulent energy
is close to what is expected from virial equilibrium. The turbulent
velocities exist primarily on small scales in the cloud and the energy
spectrum of turbulent fluctuations, $E(k) \propto k^{-\beta}$, is
derived by various methods to have $\beta \approx 1$--2. The turbulent
dissipation rate estimated using the results of current numerical
simulations is $\dot{E}_{\rm diss} \approx \dot{E}_{\rm input}$. 

This study reveals that stellar feedback is a significant factor in
the evolution of the L1551 cloud. 
\end{abstract}

\keywords{stars: formation --- ISM: clouds --- stars: pre-main
  sequence --- ISM: individual(\objectname{L1551}) --- ISM: structure
  --- radio lines: ISM --- techniques: interferometric } 

\section{Introduction}
Low-mass stars form from dense, quiescent cores embedded within
turbulent molecular clouds \citep{mye83a,bei86,lar81}. 
Following a rapid inside-out gravitational collapse phase
\citep{shu77}, embedded sources accrete material from their
surroundings through a disk while bipolar flows interact with the
progenitor cloud \citep{har95,ric00,reip01}.
Young stars become optically visible and accretion diminishes over
time, as does the mass of the circumstellar disk
\citep{har98,lad91}. The luminosities of the pre-main sequence stars
evolve as they contract and eventually settle onto the main sequence
\citep{hay61}. This succession in the formation of low-mass stars is
fairly certain, but the theoretical context in which these stages fit
is still debated.

As first pointed out by \cite{zuc74}, the molecular mass in the
Galaxy, $\sim 2 \times 10^9$\msun\ \citep{blo86}, compared
to the star formation rate, $\sim 4$\,\msun\ yr$^{-1}$ \citep{pra95},
indicates that star formation is a globally inefficient process
\cite[see also][]{kru07}. Another universal property of star formation
is the distribution of stellar masses produced by a star-forming
cloud, or the initial mass function \citep[IMF;][]{mey00}. Any
successful theory of star formation will explain these general
properties of star formation in a way consistent with the above
scenario.

In the standard paradigm for low-mass star formation \citep{shu87},
magnetically sub-critical cores evolve into protostars over timescales
of $\sim 10$\,Myr \citep{cio94}. Upon the formation of a protostar,
outflows limit the amount of material available for accretion and
disperse the parent cloud. Thus feedback from young sources play a
critical role in determining the final stellar masses and overall star
formation efficiency. The copious energies of molecular outflows
\citep{bac96}, the large spatial extent 
of Herbig-Haro flows \citep{reip01}, and the existence of remnant
cavities in star-forming clouds \citep{qui05,cun06} all lend credence
to this picture. 
However, magnetic field observations do not show strong
evidence for magnetically sub-critical cores \citep{cru99}. 

In another description of low-mass star formation, ordered magnetic
fields and stellar feedback are not critical elements of the
process. Rather star formation occurs on the order of a dynamical
timescale in regions of converging turbulent velocity fields
\citep{bal07,elm00}. The inefficiency of star formation and the IMF
arise solely from the properties of interstellar turbulence in this
framework \citep{vaz05,pad01,pad02}. Turbulence in star forming
regions is difficult to characterize from observational data alone
\citep{elm04} and support for this theory relies heavily upon computer
simulation.

The aim of this work is to gain insight into the relative importance
of different mechanisms involved in the star formation process by
doing a detailed case study of a star-forming region. The Lynds dark
cloud L1551 \citep{lyn62} is an ideal subject for this kind of study
since it is a relatively isolated, nearby ($\sim 160$\,pc),
star-forming cloud with a wealth of activity that encompasses all
phenomena known to be associated with low-mass star formation
including; a population of pre-main sequence stars~\citep{ken95},
multiple outflows~\citep{sne81,mor88,mor95,pou91}, jets~\citep{mun90},
winds~\citep{wel00,gio00}, an abundance of shocked gas~\citep{dev99},
and reflection nebulosity~\citep{dra85}. This article is based on the
work by \cite{thesis}, and summarizes some of the conclusions therein
while elaborating on some of the themes via new data and analyses.

Our new observations are presented in \S\,\ref{sObs} and contribute
significantly to the numerous observations in this region. Data
pertaining to the young stars around L1551 give insights into the star
formation history and the cloud lifetime, and Spitzer IRAC observations of
the dense core \core\ add to the understanding of the highest column
regions of L1551 in \S\,\ref{sStellar}. The general properties of the
L1551 cloud are then deduced in \S\,\ref{sL1551} using dust extinction
and \thco\ emission. 

The high-resolution CO isotopologue maps are shown in
\S\,\ref{sMolGas} where the \thco\ emission proves to be a good tracer
of stellar feedback in L1551. The nature of this feedback is explored
in \S\,\ref{MGasOutflows} where it is found that outflow in L1551 is
both stirring the ambient gas as well as excavating mass from the
cloud. The \ceo\ emission traces the mass distribution in L1551 better
than the \thco, and these data are used to probe the turbulent
velocity field in the cloud in \S\,\ref{MGasTurb}. The star formation
history in L1551 is discussed in \S\,\ref{sDisc}, and the main
conclusions of the article are summarized in \S\,\ref{sConc}.

\section{Observations and Reductions} \label{sObs}
\subsection{Wide-Field BIMA Mosaic} \label{MGasMos}
\begin{figure}[!b]
\centerline{\includegraphics[angle=270,width=3.3in]{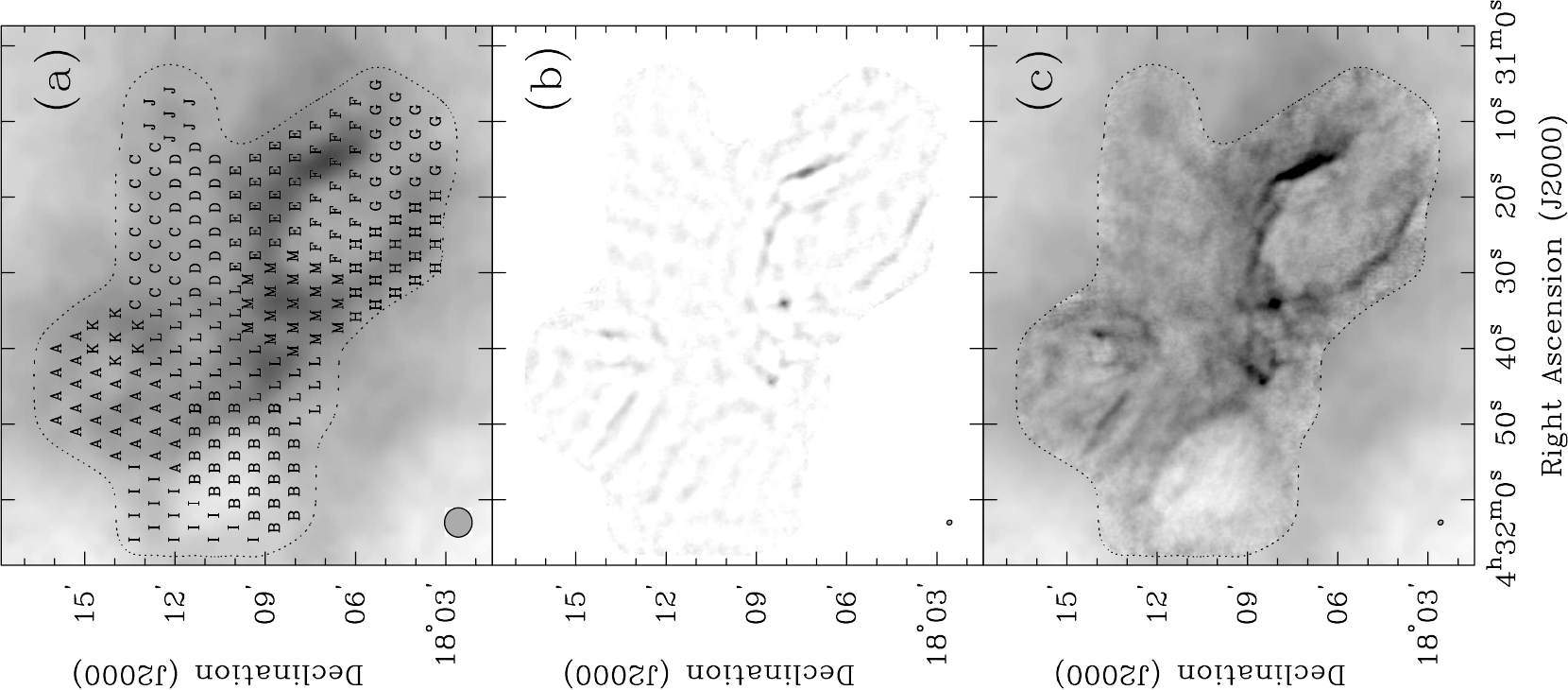}}
\caption{({\it
  a}) Velocity integrated \thco$(1-0)$ emission from the 
  Kitt Peak 12\,m showing the pointings from each sub-mosaic
  A--M. ({\it b}) The integrated emission from the interferometer
  data, and ({\it c}) the combined map as described in
  \S\,\ref{MGasComb}. The gray scale is transferred linearly between 0
  and 15\,K\,\kms\ for all images. \label{CombFig}} 
\end{figure}

A 250 pointing interferometric mosaic conducted with the
BIMA\footnote{The
Berkeley-Illinois-Maryland Association array was, at the time of
these observations, operated by the University of California,
Berkeley, the University of Illinois, and the University of Maryland
with support from the National Science Foundation. The BIMA array has
now been combined with the Owens Valley Radio Observatory's millimeter 
interferometer to create the Combined Array for Research in
Millimeter-Wave Astronomy, or CARMA.}
interferometer covers $\sim 140$ square arcminutes of the L1551
cloud. The mosaic is designed to image the regions of
highest \thco\ emission as seen in the single-dish maps of
\cite{swi05} including the region of \core. The full mosaic is broken
into 13 
sub-mosaics, labeled A--M, each consisting of roughly 20
pointings. Each sub-mosaic track was run for a full night in both the
C and D array configurations while follow-up scheduling was directed
to create nearly constant sensitivity across the
map. Table~5.1 in \cite{thesis} summarizes the observations which span
7 years with the vast majority of tracks run between 2001 and
2004. Figure~\ref{CombFig}({\it a}) displays the pointing centers for the
sub-mosaics superposed on velocity integrated \thco\ from \cite{swi05}. 

The pointing coordinates in each sub-mosaic are defined in
terms of right ascension and declination offsets from the most central
pointing in the sub-mosaic. The offsets are chosen such that the
pointing centers make a grid of equilateral triangles with each side
equal to the Nyquist sampling angle, $\Theta_{\rm N} = 46$\arcsec. The
exception is Mosaic M where both the right ascension and declination
offsets are integer multiples of $\Theta_{\rm N}$.

The mosaic data were taken in correlator mode 6~\citep{wel96} with the
two high-resolution windows containing 256 channels set to 12.5\,MHz
widths centered on the transition frequencies of \ceo\
(109.78216\,GHz) and \thco\ (110.20135\,GHz) in the upper side
band. This translates to a velocity resolution of 0.13\,\kms\ in these
spectral windows. 

Each sub-mosaic track cycled through all pointings with 23\,s
integrations performed in groups of 1, 2 or 3 at each pointing center
depending on mosaic size in consideration of proper phase
calibration. Integrations at all pointings were performed between
phase calibration for tracks with 20 pointings or less. For the larger
mosaics (A, B, and L), two calibration integrations were 
embedded in the mosaic to ensure phase tracking. In our final data,
the \uv\ sampling is complete over the entire map from $2\lesssim
\sqrt{u^2+v^2} \lesssim 25$\,k$\lambda$. The observations were 
phase calibrated using nearby quasars 0530+135 and 0429+113, both
$\gtrsim 1$\,Jy sources at these observing frequencies throughout the
program.

The flux scale was calibrated using W3(OH) as the primary calibrator.
Saturn or Uranus was used as a secondary calibrator, and Jupiter was used 
occasionally for D configuration observations. 
The flux scale in the final maps is estimated to be accurate at the
$\sim 15$\% level.

Anomalous amplitudes and phases were flagged from the data, and no
shadowed data were used. Line-length and gain calibrations were
applied, and then all data were Fourier transformed simultaneously
using MIRIAD~\citep{sau95}. Natural weighting was used to maximize
signal to noise. Figure~\ref{CombFig}({\it b}) shows the positive
values of the velocity integrated emission from the Fourier
transform---the ``dirty'' map. 

The maps were then deconvolved with a Steer-Dewdney-Ito CLEAN
algorithm~\citep{ste84} using a low loop gain of 0.05 and a maximal
number of iterations so that all flux above the 1\,$\sigma$ level was
transferred to CLEAN components. The clean image was then restored
with a bivariate Gaussian clean beam having FWHM dimensions of
$10.\!\arcsec3\times8.\!\arcsec7$ and position angle of $29^\circ$.

\subsection{Combining Single-Dish and Interferometer Data}
\label{MGasComb}  
A single pointing of an interferometer cannot measure visibilities
where $\sqrt{u^2+v^2} < D$. Observations using multiple pointings
allow the \uv\ plane to be sampled slightly inside this limit due to
the \cite{eke79} effect, but intensity variations on scales
larger than the primary beam cannot be recovered with
interferometry alone [compare Figure~\ref{CombFig}({\it a}) and ({\it 
b})]. Therefore, when mapping extended objects with interferometry the
addition of single dish data is essential in reconstructing accurate
depictions of the sky brightness distribution
\citep[see][]{hol99}. 

Approximately 6\% of the total \thco\ flux is recovered with our
cleaned BIMA mosaic. The flux recovery is greatest in the line wings,
reaching as high as $\sim 35$\%, and steadily decreasing to
effectively zero in the line core. The flux recovery in the \ceo\ map
is on the level of a few percent.

The flux density scales of the single-dish and interferometric
datasets were compared in the region of \uv\ overlap in the manner 
described by \cite{sta99}. The flux density scale for the Kitt Peak
12\,m was consistent with the nominal scale of
33\,Jy\,K$^{-1}$~\citep{hel03}, and a flux scaling factor of unity is
assumed for all data in the combination.

The data are combined using a linear combination of the two datasets
followed by a maximum entropy deconvolution \citep{sta99,hol99}. The
regridded single-dish map, seen in velocity integrated \thco\ 
intensity in Figure~\ref{CombFig}({\it a}), is first combined with the
dirty BIMA mosaic using Equation~4 in \cite{sta99}. The ratio of beam 
areas, $\alpha = 0.029$.

The maximum entropy deconvolution was performed using MIRIAD. An rms
factor of 1.2 was used to account for the discrepancy between the
theoretical and the real rms in the  
images. The regridded single-dish image served as the default image,
and the total flux of the output was forced to agree with the total
flux in the default image. Without using the single-dish
image as the default the total flux in the final image was
overestimated by nearly a factor of 2.

Convergence was achieved in all 64 image planes in $\lesssim 15$
iterations. There are no significant artifacts from the deconvolution
and the overall image quality is very good. The deconvolution
residuals contain minimal structure with a consistent rms from channel
to channel, evidence of a successful deconvolution. The regridded
single-dish map, the interferometer map, and the deconvolved
combined map integrated over the \thco\ line profile can be seen in 
Figure~\ref{CombFig}.

\subsection{Deep Near Infrared Imaging and A Large Scale Extinction Map} 
\label{L1551Extinct} 
Deep near infrared (NIR) imaging was obtained on the night of 2004 November
11 to supplement data from the 2MASS point source
catalog~\citep{sku06} in the L1551
region. The images were obtained with FLAMINGOS on the Kitt Peak 2.1\,m
telescope. FLAMINGOS is a combination wide-field, near infrared imager
and multi-object spectrometer built by the University of
Florida~\citep{els03}. In 
imaging mode, the $2048\times2048$ science grade Hawaii-II HgCdTe
array has a plate scale of 0.\!\arcsec606\,pixel$^{-1}$ and a
20.\!\arcmin7 field of view.

Three sets of exposures were taken through the $J$, $H$, and $K$-band
filters with total exposure times of 720, 720, and 875\,s consisting
of 15 45\,s exposures in $J$ and $H$, and 25 35\,s exposures in
$K$. These data were reduced using the LONGLEGS~\citep{rom06}
reduction procedure.

$J$, $H$, and $K_{\rm s}$ band photometric data from the
2MASS point source catalog were selected within a
$4^\circ$ radius of L1551 using {\it VizieR}~\citep{och00}.
Aperture photometry of 171 FLAMINGOS point sources matched with the
2MASS catalog produced zero points for the FLAMINGOS data to an 
accuracy of $\sim 0.\!\!^{\rm m}25$. The limiting magnitudes of our
FLAMINGOS data are 22.3 in $J$, 21.9 in $H$, and 21.4 in $K_{\rm s}$.

The NICER method~\citep{lom01} is an optimized technique for
converting NIR color excess of background  
stars into an $A_V$ map. Using conversion factors from the
literature~\citep{har03,rie85}, extinction values were assigned to
gridded $1.\!^\prime2$ pixels based on a weighted mean of measured
extinctions from stars falling within a $2.\!^\prime9$ FWHM Gaussian
beam centered on 
the pixel coordinates. The weighting incorporates errors in
photometry, errors due to intrinsic variations of stellar colors
determined from a control field located at
$4^{\rm h}20^{\rm m}00^{\rm s},
17^{\circ}00^\prime00^{\prime\prime}\,(J2000)$,  
and the stellar position relative to the pixel center. 

\begin{figure}[!b]
\centerline{\includegraphics[angle=270,width=3.3in]{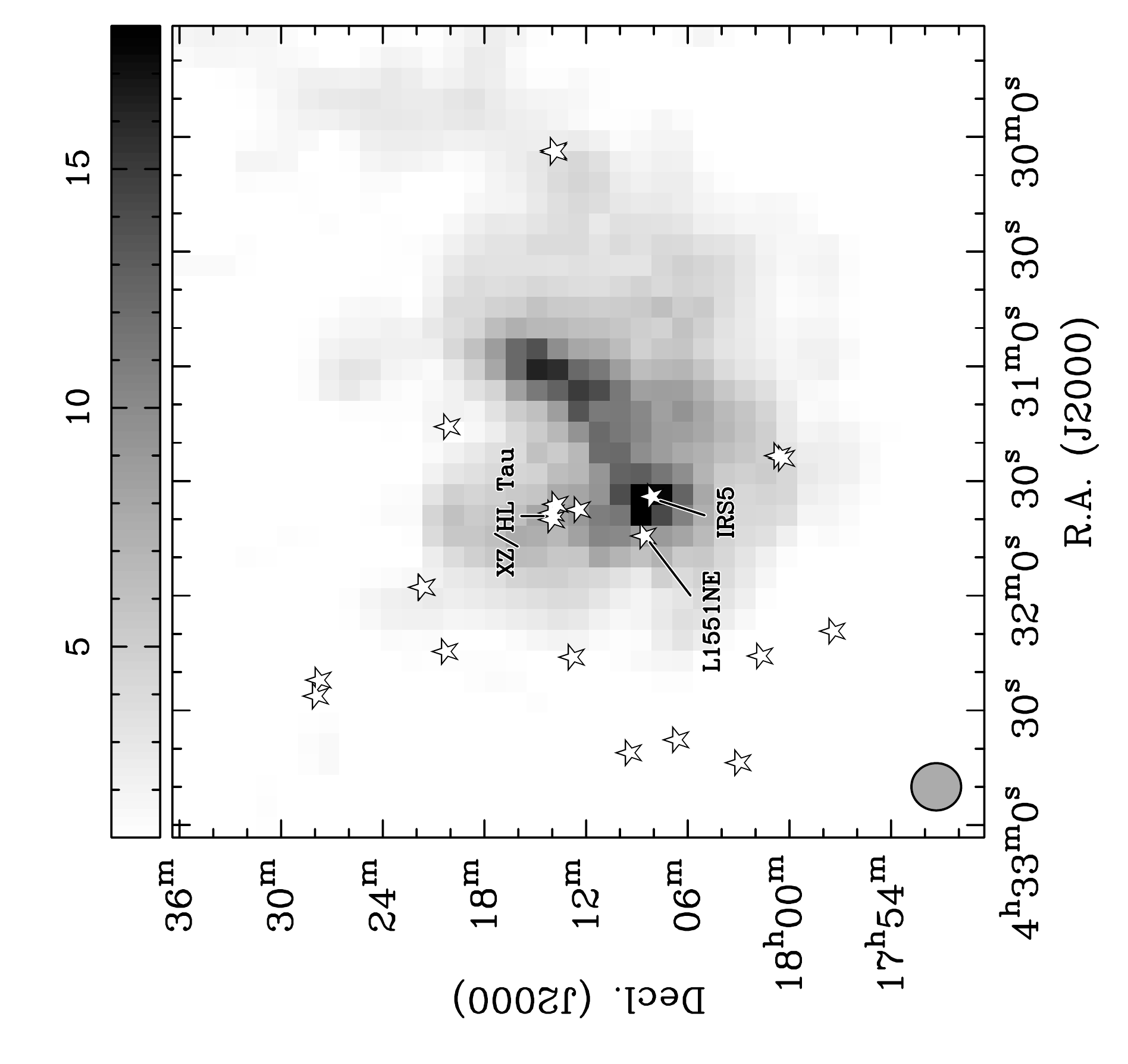}}
\caption{Map of the measured dust extinction from the L1551 cloud in
  units of $A_V$. Known pre-main sequence stars are overlaid for
  reference, and the $2^\prime\!.9$ Gaussian smoothing kernel is shown
  at the bottom left. \label{avmap}}
\end{figure}

Figure~\ref{avmap} shows the extinction map of L1551 with a dynamic
range of $0^{\rm m}\!\!.5 \lesssim A_V \lesssim 18^{\rm m}$ and a mean 
$\langle A_V \rangle \approx 6$ taken over the central
$20^\prime\times 20^\prime$ of the cloud.
The rms in the map varies from $0^{\rm m}\!\!.6$ in the part of the
map with only 2MASS data, to $0^{\rm m}\!\!.3$ in the outskirts of the
Kitt Peak 2.1\,m data, to more than $1^{\rm m}$ in the inner
regions where there are a paucity of background stars. The errors are
skewed toward higher extinction values in regions of high obscuration.

\subsection{Spitzer IRAC Imaging}
A single Infrared Array Camera \citep[IRAC;][]{faz04} footprint
centered on the position of peak \nht\ emission from
\core, $4^{\rm h}31^{\rm m}09^{\rm s}\!.9$,
$+18^\circ12^\prime41^{\prime\prime}\,\,(J2000)$~\citep{swi05}, 
was imaged with the Spitzer Space Telescope on 23 March 2006. 
In 8\,min of total on source integration time, a $3\,\sigma$ point
source sensitivity of roughly 2, 4, 10, and 10\,$\mu$Jy was achieved
in the 3.6, 4.5, 5.8 and 8.0\,$\mu$m bands, respectively. 
The post-BCD\footnote{see {\tt
http://ssc.spitzer.caltech.edu/postbcd}} data were inspected and
found to be free from artifacts in the first three bands. Small, but
noticeable ($\sim 1\,\sigma$ peak) decrements in the floor value of
several north-south columns produce the ``jail bar effect'' in our
8\,$\mu$m image. Aperture photometry provided point source flux
measurements (DN\,s$^{-1}$) using an aperture radius of 4 pixels, or
$4^{\prime\prime}\!\!.8$. Zeropoint values of 280.9, 179.9, 115.0, and
64.13\,Jy in the 4 respective IRAC bands convert these fluxes into
magnitudes.

\section{The Stellar Component} \label{sStellar}
\begin{figure}[!b]
\centerline{\includegraphics[angle=270,width=3.3in]{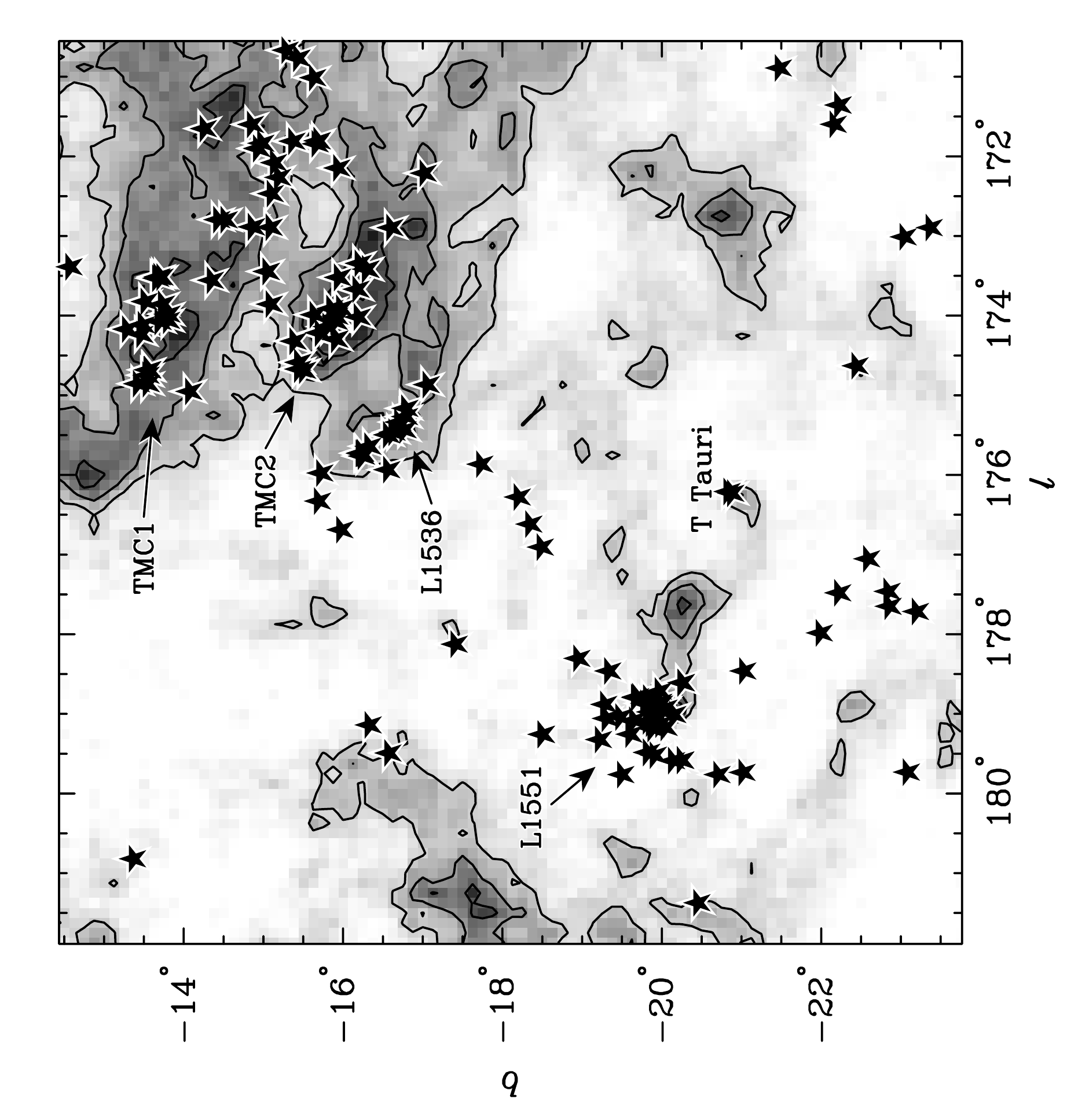}}
\caption{Southern part of the Taurus molecular complex in CO$(1-0)$
  emission shown in gray scale (0--25\,K\,km\,s$^{-1}$) and contours
  (6, 12 and 18\,K\,km\,s$^{-1}$)~\citep{dam01} with the
  positions of pre-main sequence stars ({\it star symbols})
  overlaid~\citep[][also see Table~\ref{PMStable}]{pal02}. L1551 
  is seen as a relatively isolated molecular clump coincident with an
  association of young stars. The positions of well-known star-forming
  clouds TMC\,1, TMC\,2, and L1536 as well as T\,Tau are labeled.
  \label{Taurus}} 
\end{figure}

Figure~\ref{Taurus} shows the velocity integrated CO$(1-0)$ emission
from the southern end of the Taurus molecular
complex~\citep{dam01}. To the north, the majority of young stars in
Taurus are being born along high density
filaments~\citep{oni98,har02}. 
L1551 lies further out of the Galactic plane at $l \approx
179^\circ$, and appears as an isolated molecular clump at a distance
of $160\pm20$\,pc~\citep{ber99,sne81}. 

There is an association of pre-main sequence (PMS) stars centered near
the peak CO column density of L1551. Table~\ref{PMStable} displays 70
PMS stars within a $4^\circ$ radius of L1551. There have been numerous
searches for PMS stars in this
region~\citep{her95,fei87,gom92,bri98,bri02,luh00} and it is expected
that all PMS stars within the central square degree have been
identified. 
Many searches for companion stars have also been conducted in
this region sensitive to separations between 1 and
2000\,AU~\citep{koh98,sim95,ghe93,lei93,rei93}, and a large
fraction of companion stars are presumed to be known.

\subsection{Selection} \label{L1551Select}
Stars with a high probability of having formed from the L1551 cloud
are chosen from the list in Table~\ref{PMStable} according to
their proper motions and radial velocities where possible, as well as
the projected spatial distribution.

The 34 proper motion entries listed in Table~\ref{PMStable} have
typical measurement errors of 0.5\,arcseconds per
century~\citep{duc05} translating to a $\sim 5$\,\kms\ error on the
two-dimensional space motions. The mean proper motion for this sample
is $\langle\Delta\alpha, \Delta\delta\rangle =
(0.71\pm0.1,-1.5\pm0.1)$\,arcseconds per century. There are three
outliers in the distribution of proper motions that lie beyond
15\,\kms\ of the mean---RX\,J0433.7+1823, RX\,J0430.8+2113, and
RX\,J0444.4+1952---and are rejected from the sample. 

The distribution of 21 radial velocities have a mean of 5\,\kms\ (LSR)
and a dispersion consistent with the quoted
measurement errors for this sample, $\sim 2$\,\kms\ \citep{her95}.
The only clear outlier in this distribution is DQ\,Tau, and it is
rejected based on this criterion.

\begin{figure}[!b]
\centerline{\includegraphics[angle=90,width=3.3in]{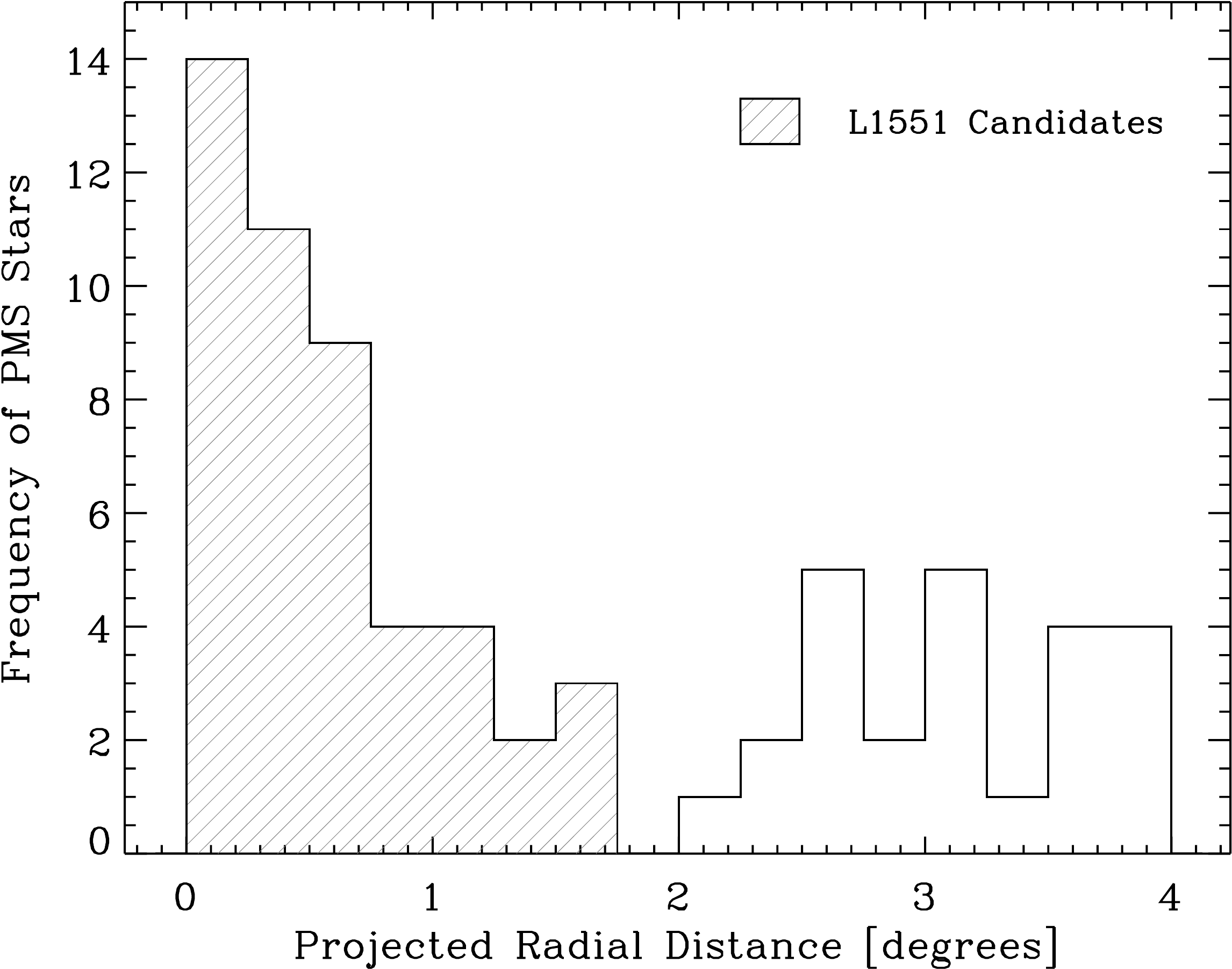}}
\caption{Histogram of the number of pre-main sequence (PMS) stars
  within $0.\!\!^\circ25$ 
  bins of projected radius from L1551. There is an absence of PMS
  stars between $1.\!\!^\circ6$--$2.\!\!^\circ1$, and this gap is used
  to separate the L1551 candidates ({\it cross-hatched area}) from PMS
  stars in the field that are less likely to have formed in
  L1551. \label{PMSprofile}}
\end{figure}

Figure~\ref{PMSprofile} shows the projected radial distribution of all
PMS stars from Table~\ref{PMStable}. The stellar density drops
smoothly to zero and then persists at a roughly constant, low level
beyond $2^\circ$. Based on this distribution, all stars beyond a
$1.\!\!^\circ6$ radius from L1551 are rejected from our sample.

Of the spatially selected PMS population, 5 have unknown spectral
types and 1 has inadequate photometry. These stars are flagged from
the sample. Main-sequence colors
\citep{ken95} are assumed for weak-line T-Tauri stars (wTTSs) and the
NIR color-color locus from \citep{mey97} is assumed for
classical T-Tauri stars (cTTSs). Seven stars in the sample have NIR
colors that are suspect or inconsistent with these assumptions
and they are marked with a ``C'' in Table~\ref{PMStable}. Six of these
seven have adequate supplementary data from the literature that are
used in further analyses, while the remaining source,
RX\,J0437.4+1851\,A, is flagged. The number of stars used in the
following analyses totals 37 (39 if the binary companions to \irsf\
and \lne\ are counted), 7 (9) of which are embedded. 
There are 29 stellar systems and 10 companions in the sample for a
companion star fraction of 34\%. This is lower than, but comparable to
the measured companion star fraction in large samples of young stars
\citep[{\eg},][]{koh98}.

\subsection{The \hr\ Diagram} \label{sHR}
The spectral types of the PMS stars are
converted to effective temperatures using \citet[][Table~A5]{ken95} for
spectral types earlier than M0 and \citet[][Table~8]{luh03b} for later
spectral types. Effective temperatures for fractional spectral types
are computed by interpolation. 
\begin{figure}
\centerline{\includegraphics[angle=90,width=3.3in]{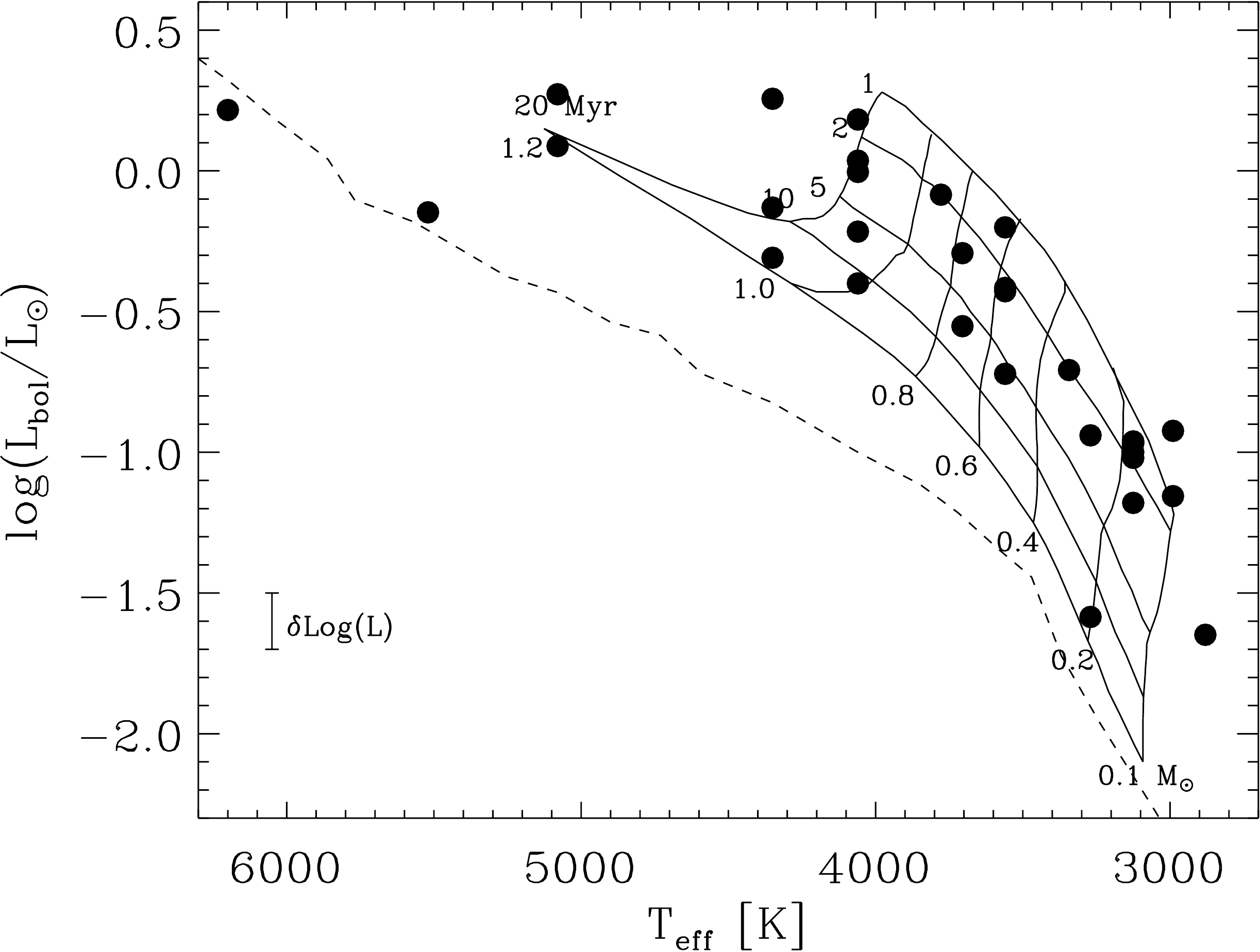}}
\caption{Hertzsprung-Russell diagram for the 30 stars that passed our
  selection criteria of \S\,\ref{L1551Select} (embedded sources are
  excluded). The assumed error in luminosity is
  shown at the bottom left (see \S\,\ref{sHR}). Pre-main sequence
  evolutionary tracks 
  for masses between 0.1--1.2\,\msun\ and ages between
  1--20\,Myr are overlaid \citep{bar98,bar02}. The two stars to the
  upper left, TAP\,51 and HBC\,407, lie near the main sequence ({\it
  dashed line}) and are likely
  interlopers. \label{tracks}}     
\end{figure}

The extinction toward each source is derived by tracing the reddening
line back to either the cTTS locus or the main sequence, otherwise
literature values are used.
The NIR fluxes were then
corrected according to the models of \citet{rie85}. Bolometric
luminosities are computed by applying a bolometric correction to the
extinction corrected $J$ band magnitude, $J_{\rm
c}$. \citet[][Table~A5]{ken95} is used for spectral types earlier
than M0, \citet[][Table~2]{bes91} for spectral types M0 through M7 and
\citet[][Table~2]{rei01} for spectral types M8 through M9.5.

Figure~\ref{tracks} displays the selected sample of PMS stars, excluding
embedded sources, on the theoretical \hr\ diagram. A subset of the PMS
evolutionary tracks by \cite{bar98,bar02} are overlaid. The errors on
the data points in this 
plot are based on arguments by \cite{har01b}. 
Our assumed error of $\delta\log L = 0.16$ is somewhat conservative
since this suggested value was derived assuming all stars to have
unresolved companions while it is expected that most, if not all,
companion stars in our final sample are resolved. 
No error is assumed on the effective temperatures since, in
conjunction with the bolometric correction, these errors will have a
small effect on the derived ages.

\subsubsection{Derived Stellar Properties}
Table~\ref{PMSprops} displays the ages and masses of the L1551
population derived from the evolutionary models of \cite{bar98,bar02}
along with the type, class \citep{lad87,ada87},
and the derived $A_V$ and $J_{\rm c}$ values. The errors in the
determined age are calculated by re-deriving these quantities for $\log L \pm
\delta\log L$. The errors in mass are estimated to be $\lesssim
25$\% for a half spectral type uncertainty. Two stars fall very near
the zero-age main sequence---HBC\,407, and TAP\,51---and are
heretofore rejected based on the likelihood that they are interloping
main-sequence stars.

\begin{figure}[!b]
\centerline{\includegraphics[angle=90,width=3.3in]{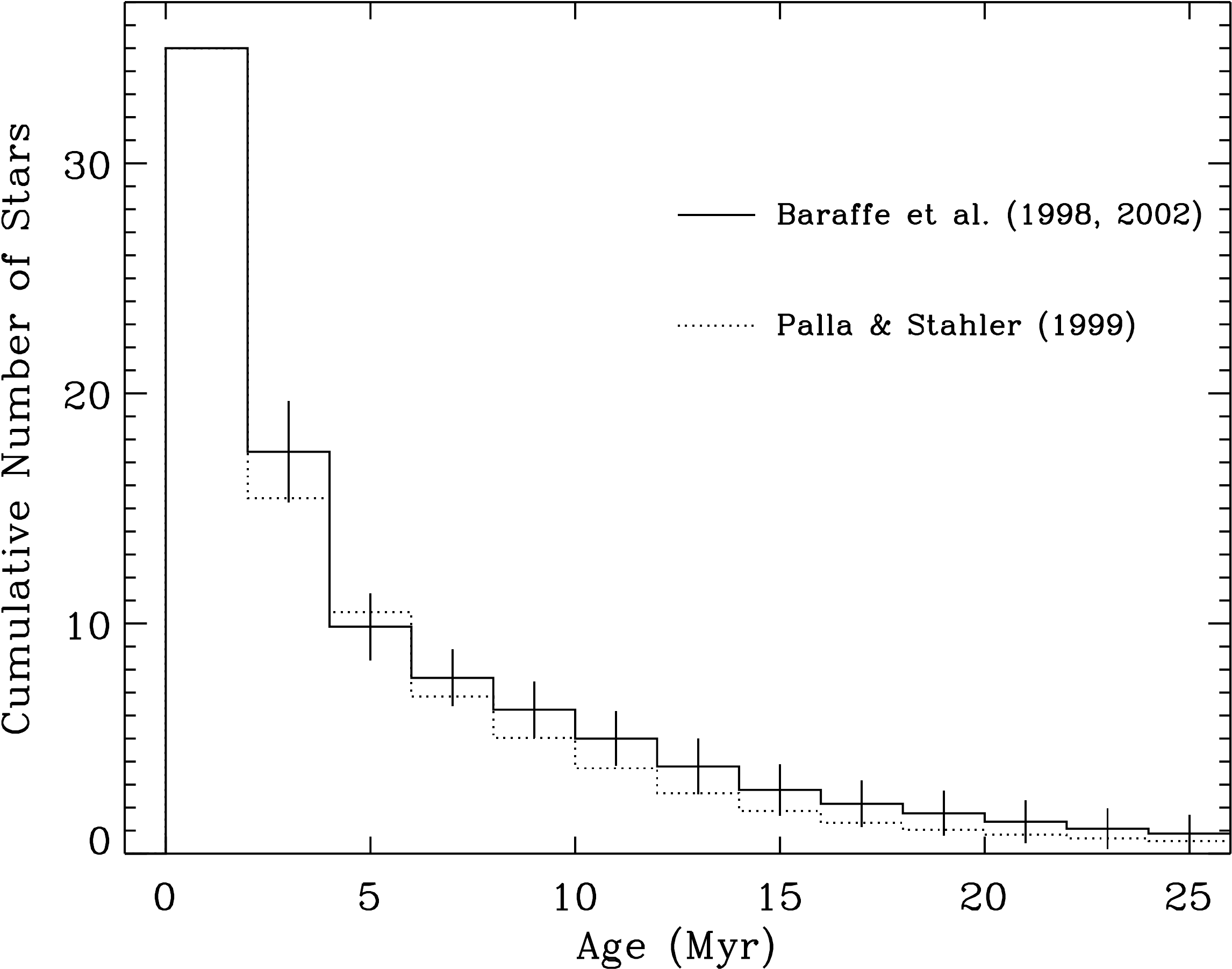}}
\caption{Cumulative histogram of pre-main sequence ages for
  the L1551 association 
  derived from the \hr\ diagram. Each 2\,Myr bin represents the number
  of stars with ages greater than or equal to the bin age. The solid
  line represents the ages obtained with models of \cite{bar98,bar02}
  and the dotted line represents the ages obtained with the \cite{pal99}
  models. \label{agehist}} 
\end{figure}

Figure~\ref{agehist} shows the cumulative age distribution of pre-main 
sequence stars in L1551. Each bin value represents the number of stars
associated with L1551 that have ages greater than or equal to the age
value of the bin, and the errors are determined using a Monte-Carlo
technique. There are a total of 35 stars in this sample, 28
have theoretical ages from Figure~\ref{tracks}, and the 7 embedded sources
assumed to have ages $< 1$\,Myr fall within the first bin. The total
stellar mass of this sample is $22\pm5$\,\msun. 

The majority of the selected PMS stars have ages less than
4\,Myr. However, both age distributions in Figure~\ref{agehist}
derived from independent PMS evolutionary models show
that $\sim 20$\% of the stars in this region formed more than 6\,Myr
ago. While the derived age spread in this region is larger than the
age spread typically stated for the Taurus complex as a whole
\citep[\eg,][]{bal99,har01a}, these results are consistent with
previous studies~\citep{gom92,pal02}. 

\subsection{The Spatial Distribution of the PMS Population} \label{sSpatial}
The radial distribution of PMS stars in L1551 is concentrated near the
center of the cloud, and decreases to zero at a radius of
$R_{\rm stars} = 1.\!\!^\circ6$, or 4.5\,pc in projected distance. 
The dispersion timescale for the stellar association is estimated
\begin{equation}
t_{\rm disp}^* = \frac{\pi}{2}\frac{R_{\rm stars}}{\sigma_{\rm 3D}^*}
\label{tdispstars}
\end{equation}
where $\sigma_{\rm 3D}^*$ is the three-dimensional velocity dispersion
of the stars, and the factor of $\pi/2$ accounts for a random
inclination angle of stellar motion with respect to the line of
sight. The dispersions of the proper motion and radial velocity
distributions of the L1551 association are approximately equal to the 
stated measurement errors \citep{duc05,her95}. Therefore an upper
limit on the one-dimensional velocity dispersion for the stellar
association of 2\,\kms\ results from the dispersion of radial
velocities. A lower limit of 0.3\,\kms\ is derived from the velocity
dispersion of \ceo\ emission in the cloud (\S\,\ref{MGasTurb}).  

For a one-dimensional velocity dispersion of 1\,\kms\ ($\sigma_{\rm 3D}
= \sqrt{3}\sigma_{\rm 1D}$), $t_{\rm disp}^* \approx 4$\,Myr. The
observed spatial 
distribution of stars is thus consistent with dispersion by random
motions over several million years. It is also possible that the
stars formed in their observed locations within the last $\sim
1$\,Myr, but this seems less likely given the distribution of
molecular gas in relation to the stellar positions and the high
fraction ($\sim 50$\%) of wTTSs in the association. 

The radial distribution of stars fans out to the east, and there are
significantly more stars to the east of the cloud center than the
west (see Figures~\ref{avmap} and \ref{Taurus}). \cite{mor06} argue
that L1551 is being eroded from the east by ionizing flux, perhaps
from Orion, and the PMS stellar distribution is modest support for their
conclusion that star formation is progressing predominantly from east
to west. 

Although we are only observing projected separations, the spatial
distribution of embedded stars does not appear to follow any
particular prescription such as Jeans fragmentation. All embedded
stars appear in regions of stellar activity suggesting that triggering
may be an important process \citep[{\eg},][]{yok03}. However, the fact
that the next star or stellar system in L1551 appears to be forming
from quiescent gas (\S\,\ref{sNext}) counters this notion. 

\subsection{Future Star Formation} \label{sNext}
A gravitationally bound, 2--3\,\msun\ pre-protostellar core was
recently discovered in a quiescent region to the northwest of \irsf\
\citep{swi05,mor06}. Follow up observations of this core, named \core,
confirm the existence of high volume density gas and reveal $\gtrsim
0.1$\,\kms\ infall signatures~\citep{swi06}. The gas motions related
to the redshifted self-absorption in the CS line profiles are
associated with build up of core mass, and it is probable that a star
or stellar system will form in this region in $\lesssim
1$\,Myr. However, the existence of a very low luminosity object
\citep{kau05} embedded in \core\ has not yet been ruled out. 
 
\begin{figure}
\centerline{\includegraphics[width=3.3in]{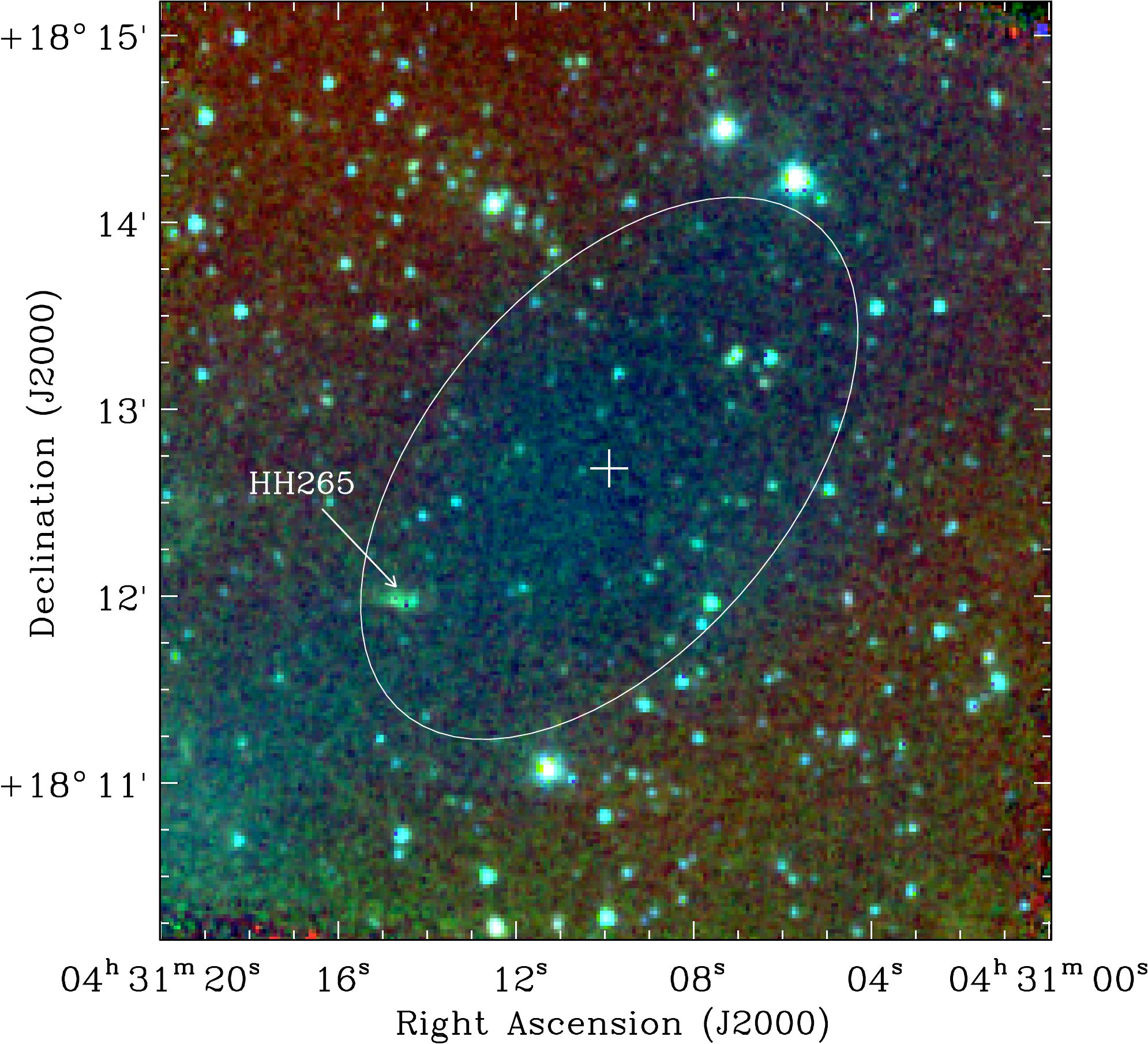}}
\caption{Spitzer IRAC composite image of \core\ with blue $=
  3.6\,\mu$m, green $= 4.5\,\mu{\rm m}+5.8\,\mu{\rm m}$, and red $=
  8.0\,\mu$m. The 
  position of peak \nht\ emission from \cite{swi05} ({\it
    white cross}) and an ellipse enclosing the highest
  column density region of the map are shown. \label{ch2}}  
\end{figure}

Figure~\ref{ch2} shows a composite image of \core\ made from all four
IRAC bands. There is a deficit of emission along the central dust lane
in the 8\,$\mu$m band due to the extinction of the Galactic background
at these wavelengths. In the 3.6 and 4.5\,$\mu$m bands there is excess
emission along the lane due to scattered light. There is no 
evidence for an embedded source within an arcminute of the peak \nht\
emission in our IRAC data. An upper limit of $L_{\rm bol} <
0.005$\,\lsun\ is set from 
the sensitivity of our 8\,$\mu$m IRAC image using a bolometric
temperature of 100\,K.

Two 24\,$\mu$m sources appear near \core\ in MIPS archival data
(Campaign ID: 713). The brighter source has a flux $\sim 1$\,Jy and is
located at $4^{\rm h}31^{\rm m}05^{\rm
s}\!\!.7$,$+18^\circ13^\prime21^{\prime\prime}\,\,(J2000)$. This 
source has no counterpart in any of the IRAC bands and shows
a systematic $0^{\prime\prime}\!\!.22$\,s$^{-1}$ motion relative to
another point source in the MIPS field. It is likely this source is an
asteroid. The weaker source is located at $4^{\rm h}31^{\rm m}07^{\rm
s}$,$+18^\circ13^\prime17^{\prime\prime}\,\,(J2000)$  and is 
coincident with an extended source apparent in all IRAC bands. This
object may be either multiple sources along the line of sight or a
background galaxy.

The 3.6 and 4.5\,$\mu$m images are used to estimate the total visual 
extinction inside and outside the central $\sim 2^\prime \times
3^\prime\!.5$ region of \core\ shown in Figure~\ref{ch2}. The mean
$[3.6]-[4.5]$ color of stars inside this oval translates to $\langle  
A_{V,in}\rangle  = 42^{\rm m}\pm 4^{\rm m}$. The colors of background
stars outside the oval give $\langle A_{V,out}\rangle =
19^{\rm m} \pm 3^{\rm m}$, consistent with the NIR measurements in
this region.

A low-mass star or stellar system is likely to form within the high
column dust lane extending northwest from \irsf, and these new Spitzer
data show that a low luminosity protostar has yet to form in the
\core\ region. Given the limited extent of the dust lane (see
Figure~\ref{avmap}), it is unlikely that a significant number of stars
will form in the future. However, further study is needed to determine
the gravitational stability of the gas northwest of \core.
\bigskip
\bigskip
\bigskip

\section{General Properties of the L1551 Cloud} \label{sL1551}
\subsection{Mass and Abundances} \label{sL1551Mass}
The visual extinction shown in Figure~\ref{avmap} is converted to
total hydrogen column using \cite{boh78} and an extinction curve
slope, $R_V = 3.1$, giving 
$N$(H$_{\rm tot}) = 1.9\times10^{21}\,A_V$\,cm$^{-2}$. Summing over
the entire cloud yields a mass of $M_{\rm H} \approx
120$\,\msun. Correcting for a helium abundance, $Y_{\rm p} =
0.25$, the total mass of L1551 $M_{\rm tot} \approx 160$\,\msun.
This value is higher than all previous estimates of the ambient cloud
mass made using CO isotopologue emission
\citep[{\eg},][]{sne80,sto06}. CO fails to trace areas of low
extinction due to photodissociation and also suffers
saturation effects in high column regions. These effects are expected
to account for the discrepancies.

Assuming local thermodynamic equilibrium (LTE)
conditions, the optical depth of both \thco$(1-0)$ and \ceo$(1-0)$ can
be estimated over the central $20^\prime\times20^\prime$ of
L1551~\citep[][Equation~3]{swi05}. The
maximum \thco\ optical depth in L1551 is computed to be $\tau_{\rm
13,max} \sim 10$, and the mean $\langle \tau_{13} \rangle \sim 2$. The
\ceo\ emission is optically thin everywhere.

A pixel by pixel comparison of our $A_V$ map with the regridded CO
isotopologue maps gives a conversion between magnitudes of visual
extinction and total CO isotopologue column. We assume a constant
excitation temperature of 15\,K for the molecular gas, although
variations from 9 to 25\,K may exist~\citep{swi05,sto06,sne81}. There
is a linear correlation between both the $\tau$-corrected \thco\ and
the \ceo\ column depth with $A_V$ between $\sim 2$ and $10^{\rm
m}$. For $A_V \gtrsim 10^{\rm m}$ the 
scatter in the relation becomes large and the CO derived column depths
appear to saturate. Linear fits to the column depths as a function of
$A_V$ between 0 and $10^{\rm m}$ give
\begin{eqnarray}
N(^{13}{\rm CO}) = (2.5\pm0.1)\times10^{15}\,(A_V-0.8\pm0.3)\,{\rm cm}^{-2}
\label{thcoav} \\ 
N({\rm C}^{18}{\rm O}) = (2.6\pm0.2)\times10^{14}\,(A_V-2.2\pm0.4)\,{\rm
cm}^{-2}  
\label{ceoav}
\end{eqnarray}
with the 1\,$\sigma$ error estimates from the fits included explicitly.

These numbers agree well with past
studies~\citep[\eg,][]{dic78,lad94}. Using \cite{boh78}, the abundance
of \thco\ is $1.3\pm0.05\times10^{-6}$ for $A_V$ between 0.8 and
$10^{\rm m}$, and the abundance of \ceo\ is $1.4\pm0.07\times10^{-7}$
relative to the total hydrogen content. Using an excitation
temperature of 25\,K increases the abundance of the isotopologues by
$\sim 25$\%. 

\subsection{Energy} \label{L1551Energy}
\begin{figure}
\centerline{\includegraphics[angle=90,width=3.3in]{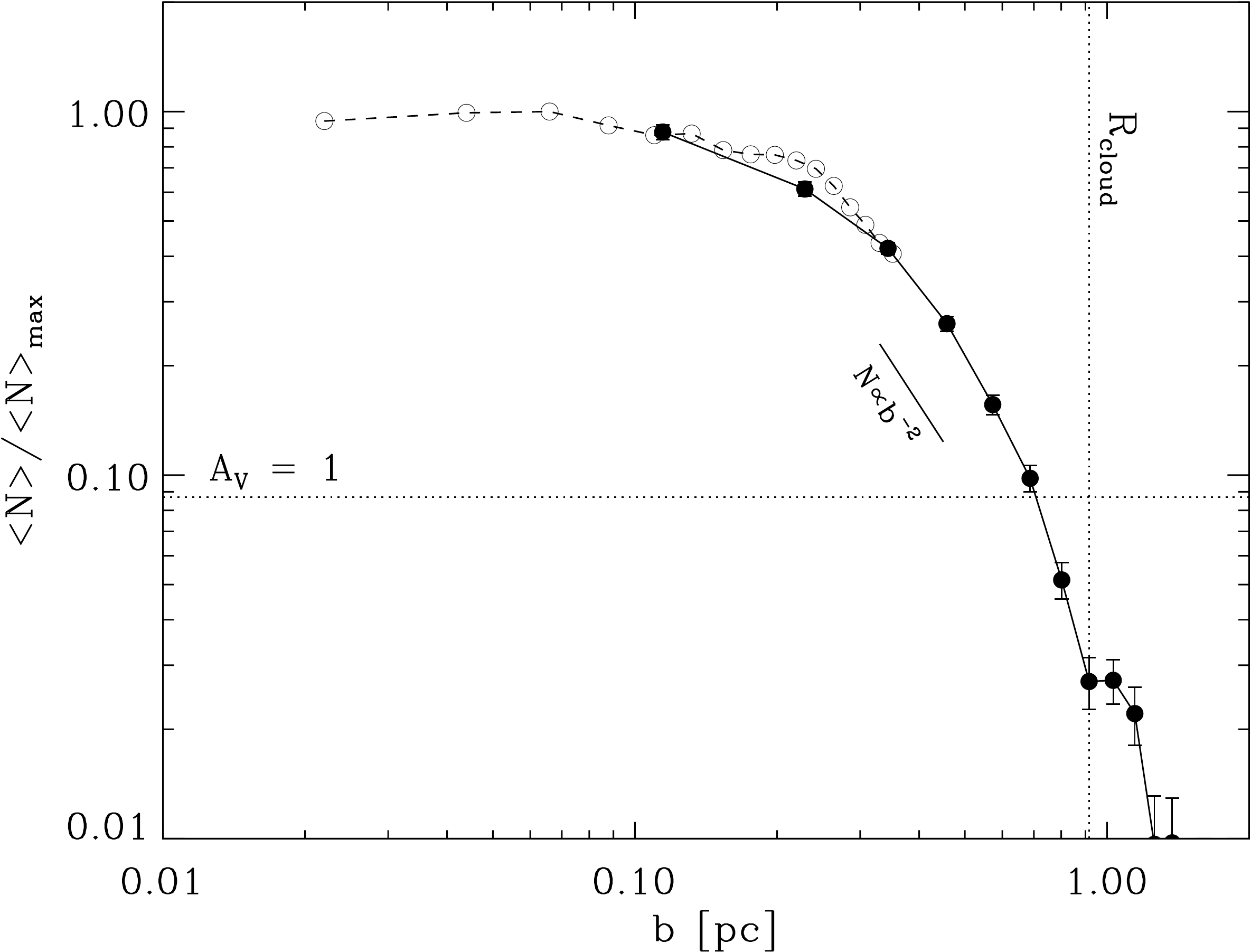}}
\caption{Normalized column density profile of the L1551 dark cloud 
  derived from both $\tau$-corrected \thco\ emission ({\it open
  circles}) and near infrared extinction ({\it filled circles}). The
  maximum averaged column 
  density is $\langle N({\rm H}_{\rm tot}) \rangle_{\rm max}
  = 2.2\times10^{22}$\,cm$^{-2}$. \label{L1551profile}}   
\end{figure}

The gravitational energy of L1551 depends on the total mass and the
distribution of matter. 
A radial profile is computed using averages of
the $\tau$-corrected \thco\ and $A_V$ maps within circular 
annuli centered on the \ceo\ intensity weighted mean position of
L1551, $4^{\rm h}31^{\rm m}24^{\rm s}$,
$+18^\circ10^\prime00^{\prime\prime}\,\,(J2000)$.
Figure~\ref{L1551profile} shows this 
column density profile from $\sim 0.02$ to 2\,pc. The error
bars for each value are determined from the rms deviation of pixel
values and the number of pixels in each annulus. The errors derived
for the \thco\ data lie within the symbols. The profile is flattened
in the inner region of L1551 and becomes progressively steeper out to
the edge of the cloud, $R_{\rm cloud} = 0.9$\,pc. 

For a density profile
\begin{equation}
\rho(r) = 
\begin{cases} 
  \rho_0  & ; \quad 0 \leq r \leq R_{\rm core} \\
  \rho_0\left(R_{\rm core}/r\right)^2 & ;
  \quad R_{\rm core} \leq r \leq R_{\rm cloud}  \\
\end{cases}
\label{brokenprofile}
\end{equation}
with $R_{\rm core} = 0.1$\,pc, $R_{\rm cloud} = 0.9$\,pc, 
and $M_{\rm tot} = 160$\,\msun, the  gravitational energy
\begin{equation}
E_{\rm grav}^{\rm core} = -\frac{3}{5}\frac{GM_{\rm core}^2}
{R_{\rm core}}\left[1+5\frac{\left(R_{\rm cloud}-R_{\rm core}\right)}
{R_{\rm core}}\right]
\label{Egravcore}
\end{equation}
is $-8.7 \times 10^{44}\,{\rm ergs}$,where $\rho_0 =1.9 \times
10^{-19}$\,g\,\cmt\ is found by normalizing the total 
mass to 160\,\msun, and $M_{\rm core} = 4/3\,\pi\rho_0R_{\rm
core}^3$. An upper limit to $E_{\rm grav}$ is given by the
gravitational energy 
of a truncated, singular isothermal sphere (SIS) with the total mass
and radius of L1551, $E_{\rm grav}^{\rm SIS} = -GM^2/R = -2.5 \times
10^{45}$\,ergs. In light of these calculations, we use a
gravitational energy of $E_{\rm grav} = -1\times10^{45}$\,ergs in
further analyses with a factor of $\lesssim 2$ uncertainty.

The turbulent velocity width derived from the composite \ceo\ spectrum 
(see \S\,\ref{MGasTurb}) is $\sigma_{\rm turb} =
0.52$\,\kms. The one-dimensional thermal width of the
molecular gas 
\begin{equation}
\sigma_{\rm therm} = \left(kT/\bar{m}\right)^{1/2}
\label{sigmatherm}
\end{equation}
is 0.23\,\kms\ for $T = 15$\,K and $\bar{m} =
2.3\,m_{\rm H}$. Adding in quadrature the turbulent and thermal
contributions gives the three-dimensional velocity dispersion of the
gas in L1551, $\sigma_{\rm v} = 0.65$\,\kms. The total kinetic energy
is thus $6.8 \times 
10^{44}$\,ergs, very near what the virial theorem predicts given our
estimation of $E_{\rm grav}$.

\subsection{Timescales and the Jeans Criterion}
\label{sL1551Stab}
Approximating L1551 as a homogeneous sphere with density $\langle n
\rangle \approx 1000$\,\cmt\ and mean particle mass $\bar{m} =
2.3\,m_{\rm H}$, the free-fall timescale $t_{\rm ff} =
\left[3\pi/(32\,G\,\rho)\right]^{1/2} = 1.1$\,Myr. The dynamical
timescale of the cloud is taken to be equal to the free-fall
timescale, $t_{\rm dyn} = t_{\rm ff}$. The sound crossing time in
L1551 is estimated as $t_{\rm cross} = 2\,R_{\rm cloud}/c_{\rm s} =
7.7$\,Myr for $c_{\rm s} = \left(kT/\bar{m}\right)^{1/2}$ and 
$T_{\rm k} = 15$\,K. This means that L1551 cannot be supported by
thermal pressure alone.  A varying temperature structure could alter
this number significantly. However, a kinetic temperature of $>
700$\,K would be needed for $t_{\rm cross} = t_{\rm ff}$.  

Another assessment of the balance between thermal pressure and gravity
is given by the Jeans criterion \citep{jea61}. For a uniform,
isothermal gas of density $\rho$, perturbations exceeding a length
scale of $\lambda_{\rm J} = \left(\pi c_{\rm s}^2/G \rho\right)^{1/2}$
will be unstable to gravitational collapse. This length scale in L1551
is $\lambda_{\rm J} \approx 0.8$\,pc---about half the cloud
diameter---meaning it is Jeans unstable. Therefore other mechanisms,
such as magnetic fields or turbulence, must contribute to the overall
gravitational stability of the cloud to prevent a global collapse.

Unfortunately, little is known about the magnetic field in L1551 (see
\S\,\ref{sDisc}). 
However in the next few sections, a closer look at
the molecular gas in L1551 gives insight into sources and dissipation
of kinetic energy that are keeping this cloud close to virial
balance.

\section{The Molecular Gas at High-Resolution} \label{sMolGas}
The ubiquity of outflow from young stellar objects and the presence of
$\sim 35$ proto- and pre-main sequence stars associated with L1551
imply a complex relationship between the stars and the gas in
L1551 over its lifetime. The interplay between young stars and the gas
from which they formed is best revealed in our fully sampled,
high-resolution images of \thco\ and \ceo\ emission. 

\subsection{Velocity Integrated Emission} \label{MGasMaps}
\begin{figure}[!b]
\centerline{\includegraphics[angle=90,width=3.3in]{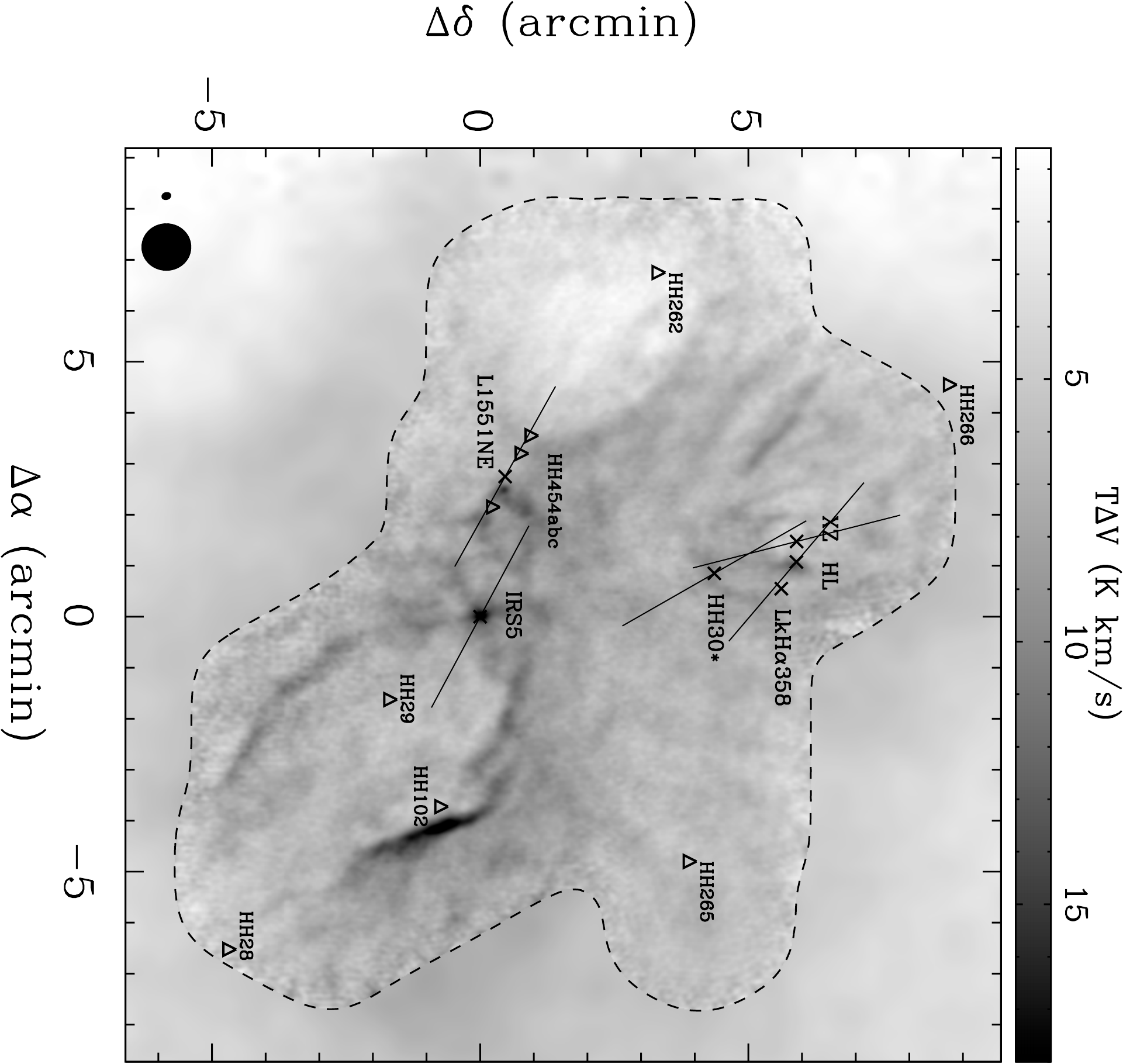}}
\caption{Velocity
  integrated \thco$(1-0)$ emission of the combined  
  interferometric and single-dish data. The image maximum is
  20.6\,K\,\kms. The axes are 
  labeled in reference to the position of \irsf, $4^{\rm h}31^{\rm
  m}34^{\rm s}\!\!.1$,
  $+18^\circ08^\prime04^{\prime\prime}\,\,(J2000)$. Sources 
  are shown as crosses and the locations of selected HH objects are
  marked with triangles. The thin lines represent the jet directions
  from the embedded sources and the resolution of the inner and outer
  regions of the map are shown at the bottom
  left. \label{MGasHires13}} 
\end{figure}
\begin{figure}
\centerline{\includegraphics[angle=90,width=3.3in]{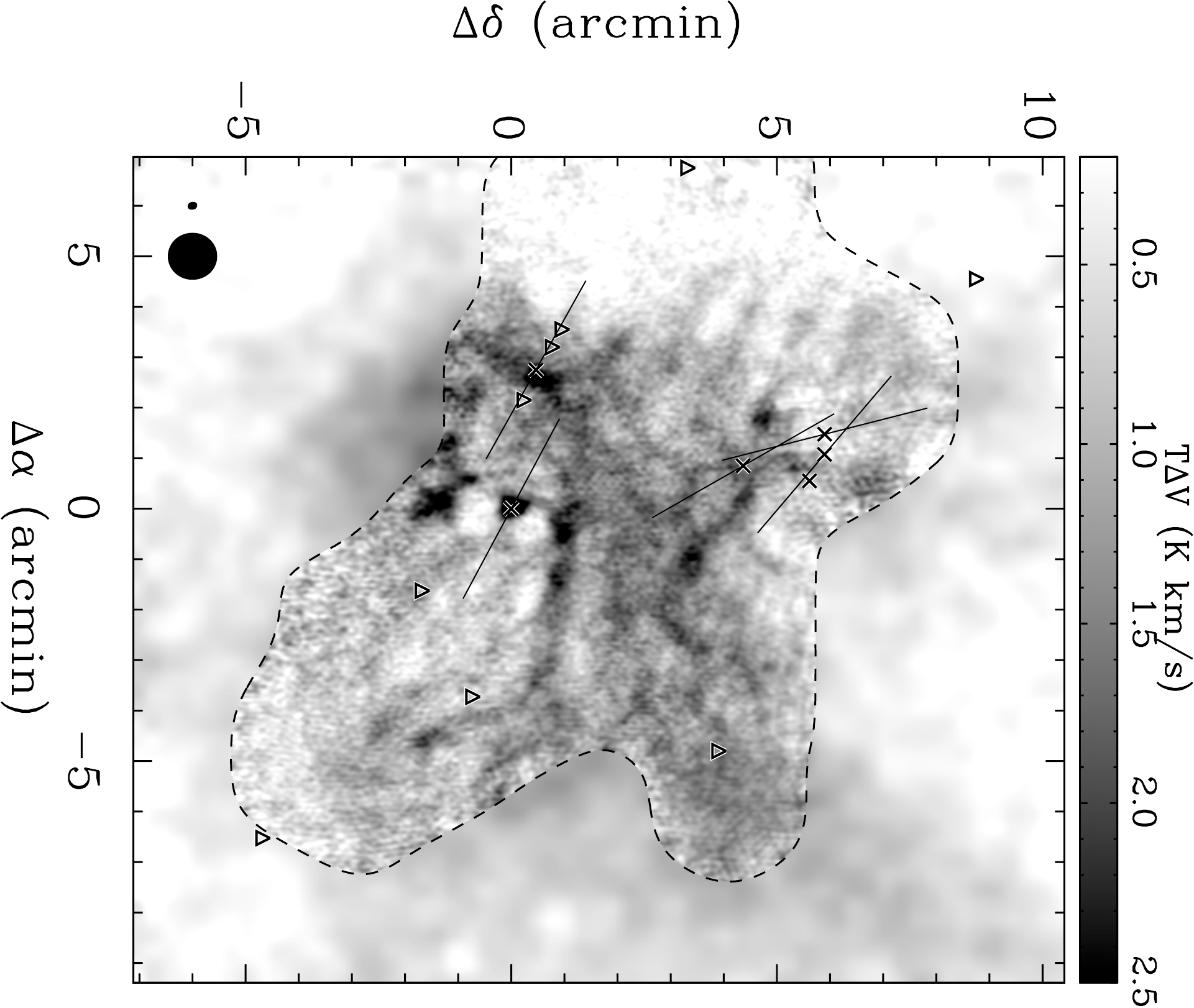}}
\caption{Same as
  Figure~\ref{MGasHires13} but for the velocity 
  integrated \ceo$(1-0)$ emission. The maximum pixel
  value in the map is 5.5\,K\,\kms\ at  
  the position of \irsf, but the linear stretch has been adjusted to
  bring out lower level features. Note the sky coverage for the \ceo\
  data is not exactly the same as the \thco\ data. \label{MGasHires18}}  
\end{figure}
\begin{figure*}
\centerline{\includegraphics[angle=90,width=6.5in]{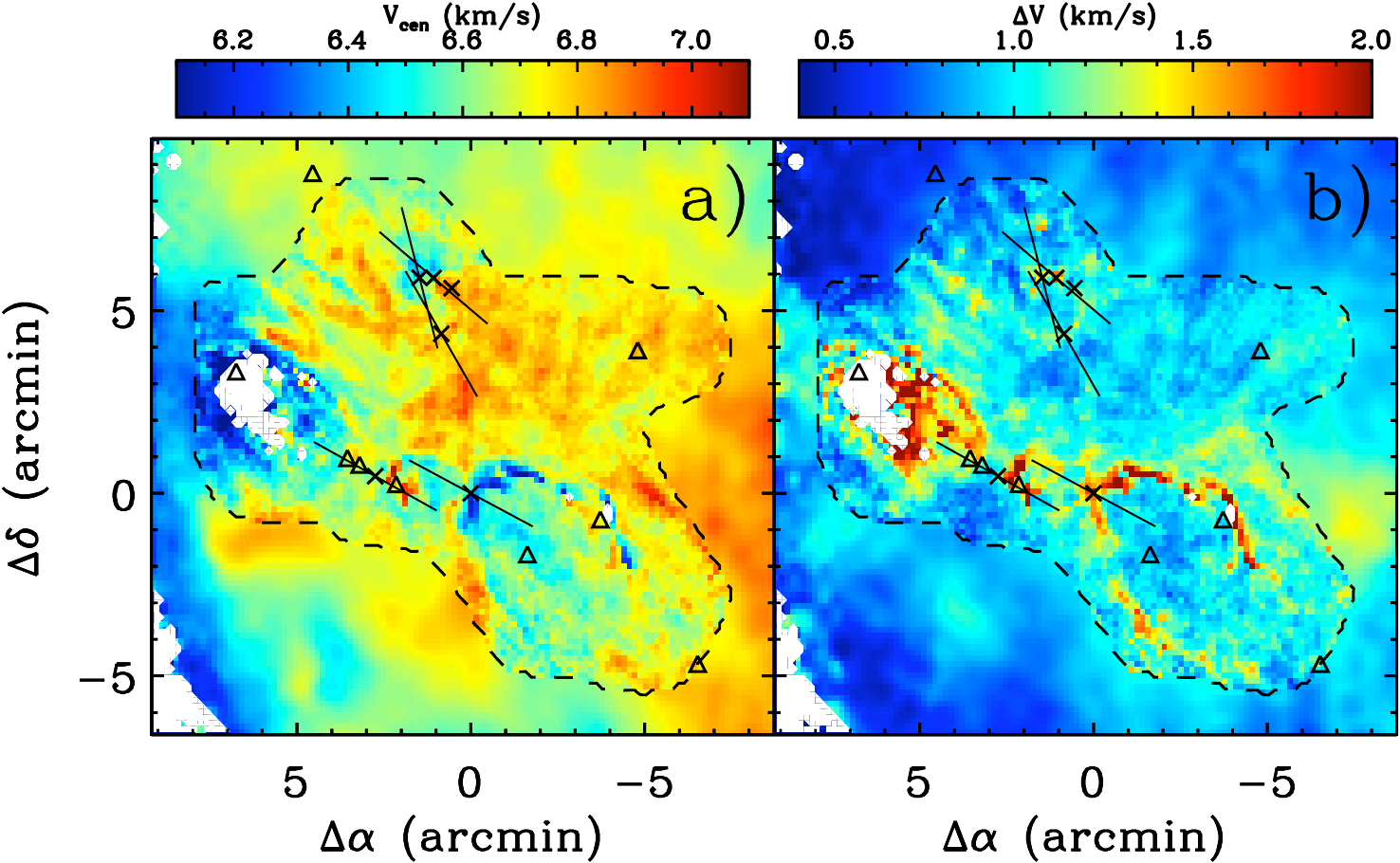}}
\caption{The first and second
  ``moment'' maps of the \thco\ emission 
  in L1551 derived from Gaussian fitting. Sources and HH objects
  from Figure~\ref{MGasHires13} are shown as well as the jet
  directions from the embedded sources. Spectra with \thco\ line peaks
  below 3\,$\sigma$ produce unreliable fits and are masked in
  white. The dashed black line encloses the high-resolution
  data. \label{MGasMoments}} 
\end{figure*}
Figures~\ref{MGasHires13} and~\ref{MGasHires18} show the velocity
integrated \thco\ and \ceo\ emission from L1551, respectively. These maps
have been created by summing $\sum T_A^*\Delta V_{\rm ch}$ over the
line profile, where $\Delta V_{\rm ch} = 0.13$\,\kms\ is the channel
width. The high-resolution data are shown within the dashed contour,
while outside this boundary there are only single-dish data. The
\ceo\ data cover a smaller range in right ascension but a slightly
larger range in declination.  

The \thco\ map shows a smooth overall
distribution of emission with clearly delineated features around
regions of known stellar energetics. Two caliper-like arcs of emission
to the southwest of \irsf\ trace the boundary of the southwest cavity
from the well known L1551\,IRS5 bipolar outflow
\citep{sne80,dra85,mor87,mor88,mor06,sto06}. Northeast of \lne, a 
sharp edge in the \thco\ map creates a parabolic shape that is
symmetric around the \lne\ jet \citep{reip00}, though the lower arm is
less pronounced than the upper. There is a deficit of \thco\ emission
within this feature that correlates with a low extinction derived
from background stars (see Figure~\ref{avmap}), and we
refer to this region henceforth as the northeast cavity. The gas
around XZ/HL\,Tau shows clear signs of disruption, likely due to jets
\citep{mun90} as well as less collimated outflow 
seen in the region \citep{kri99,cof04,wel00}. 

The \ceo\ emission is more clumpy than the \thco\ emission and follows
more closely the distribution of $A_V$. Most features seen in \ceo\
correlate with features seen in \thco, but there are a few notable
differences. The northwest region of the \thco\ map near HH\,265 is
relatively featureless, while the smooth lane of extinction that
encompasses \core\ is noticeable in \ceo. A linear feature seen in
\ceo\ extending southwest from the XZ/HL\,Tau region has no
counterpart in the \thco\ map, and the southern edge of the southwest
outflow lobe from \irsf\ is not detected in \ceo. 

\subsection{$^{13}${\rm CO} Velocity Centroids and Widths}
Figure~\ref{MGasMoments} shows the first and
second ``moment'' maps of the \thco\ emission. Gaussian
fitting of the \thco\ line profile in each independent,
$10^{\prime\prime}\times10^{\prime\prime}$ pixel produced a velocity
centroid, $V_{\rm cen}$ (left panel) and velocity full width at half maximum,
$\Delta V = 2\sqrt{2\ln2}\sigma$ (right panel) used to construct these
maps. The Gaussian shape is a decent representation of the \thco\ line
profile with $\sim 85$\% of the fits having $\chi^2$ values less than
2.

The intensity weighted mean velocity of the \thco\ emission is
6.7\,\kms. Some of the smallest values of $V_{\rm cen}$ lie along the
border of the southwest lobe of the \irsf\ flow. The diffuse emission
in the northeast cavity also has very blue central velocities likely
due to outflow (see \S\,\ref{MGasChmapSec}). The red velocities at the
western edge of the map are related to the L1551W flow
\citep{pou91,mor91,sto06}. Variations at the 0.2\,\kms\ level are
found across the map that loosely correlate with features in
Figure~\ref{MGasHires13}.

The regions where young stars are injecting energy into the cloud can
be clearly identified in the velocity width map shown in
Figure~\ref{MGasMoments}\,({\it b}). The peak line widths exceed
2\,\kms\ in a very thin ($\lesssim 1600$\,AU in some sections) rim
around the southwest lobe of \irsf\ and in the northeast cavity. High
line widths are also seen around HL/XZ\,Tau and in the L1551W flow
region. The mean \thco\ line width in L1551 is 0.96\,\kms\ with a
variance of $\sim 0.05$\,\kms.

The \thco\ velocity widths in Figure~\ref{MGasMoments}({\it b}) are a
proxy for the kinetic energy in the gas along each line of sight. The
velocity widths are only significantly above average in the regions
directly associated with the energetics from young sources. Small
spatial regions containing high line widths like the border of the
southwest lobe have small diffusion timescales and suggest that these
structures are the result of current outflow activity. Stellar feedback is the
most significant source of energy input in the cloud, and the only
source noticeable in our \thco\ data.

\subsection{The Spatial and Kinematic Distribution of $^{13}${\rm CO}}
\label{MGasChmapSec}
\begin{figure*}
\centerline{\includegraphics[height=9in]{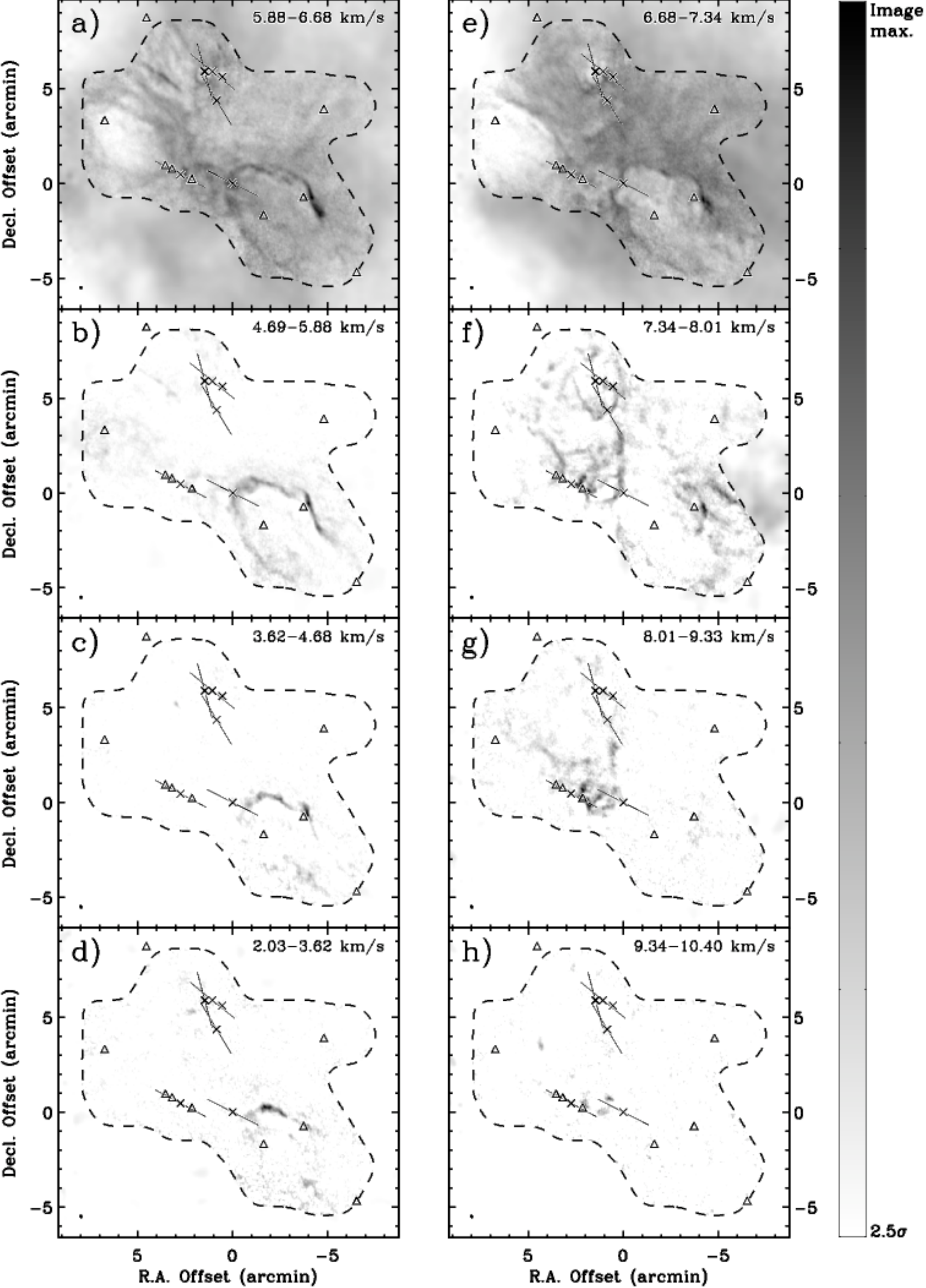}}
\caption[Channel maps of \thco\ emission]{Channel maps of
  \thco\ emission in L1551. The positions of 
  all known embedded sources ({\it crosses}), selected HH objects
  ({\it triangles}), and jet directions
  from sources ({\it thin lines}) are shown. See the text for a
  detailed description of these maps. \label{MGasChmaps}}
\end{figure*}

The nature and relevance of the features seen in
Figures~\ref{MGasHires13} and \ref{MGasMoments} are further revealed
in the \thco\ channel maps. Figure~\ref{MGasChmaps} shows the
structure of L1551 at velocity intervals chosen to highlight the
numerous features in the \thco\ data cube. The central velocity of
each map is increasingly blueward of the line center in images ({\it
a})--({\it d}) and increasingly redward in images ({\it e})--({\it
h}). Each channel map is depicted with a linear transfer function
for the gray scale shown to the right of the plot
adjusted to cover all integrated intensities from the $2.5\sigma$
level to the maximum value in each map. These values are; ({\it a})
0.40--7.4\,K\,\kms; ({\it b}) 0.33--7.6\,K\,\kms; ({\it c})
0.35--4.1\,K\,\kms; ({\it d}) 0.30--3.4\,K\,\kms; ({\it e})
0.45--6.5\,K\,\kms; ({\it f}) 0.45--3.2\,K\,\kms; ({\it g})
0.33--3.3\,K\,\kms; ({\it h}) 0.35--2.1\,K\,\kms.

\subsubsection{Synopsis of Channel Maps}
In the line core channel maps ({\it a}) and ({\it e}) much of the 
emission is optically thick, and the disruption of the ambient gas by
outflows creates the most apparent features. In channel map ({\it a}),
the northeast cavity appears as an elliptical hole with its major axis
aligned with the \lne\ jet, and the southwest cavity is outlined by a
thin parabolic arc. The limb brightened shell around XZ\,Tau
investigated in depth by \cite{wel00}, is apparent along with a few
linear structures to the east of the XZ/HL\,Tau region. 

In channel map ({\it b}), the diffuse, blueshifted emission in the
northeast cavity responsible for the small $V_{\rm cen}$ values in
this region [Figure~\ref{MGasMoments}\,({\it a})] is clearly
visible. The bright limb around the southwest lobe becomes a dominant
feature at these velocities, and the brightest emission emanates 
from a thin wall of compressed molecular gas just downstream from
HH\,102. Low level emission present near the ``mouth'' of the
southwest cavity lies along
the axis of the \lne\ jet.

Moving further into the blueshifted line wing in channel maps ({\it
c}) and ({\it d}), the bright rim of the southwest cavity
persists. The emission within the main \irsf\ flow moves downstream
but remains compact and roughly aligned with the \lne\ jet. Knots of
emission appear around the XZ/HL\,Tau region.

On the red side of the \thco\ line core shown in channel map ({\it
e}), three regions of proto-stellar disruption are visible, the
southwest cavity associated with the \irsf\ outflow, the northeast
cavity, and the XZ/HL\,Tau region. The southwest cavity is outlined by
a complex pattern, while the northeast cavity shows a sharp upper
boundary. The lower edge to the XZ\,Tau bubble is prominent
while the emission to the north of XZ\,Tau appears fragmented.

In channel map ({\it f}) the emission is broken up into a network of
thin filaments. The lower rim of the XZ\,Tau bubble is
clearly visible and extends to the position of HH\,30\,IRS. The upper limb
of the northeast cavity has a filamentary appearance at these
velocities. Northeast of \irsf, between \irsf\ and \lne, there are a
series of arcs and knots of emission. Perhaps the most interesting is
the arc that intersects the position of \irsf\ and may trace the
boundary of its red outflow lobe. The bright emission west of HH\,102
remains a prominent feature, and the L1551W flow \citep{pou91,mor91} appears
filamentary in the high resolution part of the map. 

Continuing further into the red-shifted velocities in channel maps
({\it g}) and ({\it f}) the arc intersecting \irsf\ persists but moves
gradually northeast and fades with increasing velocity. A few knots of
emission remain in the highest redshifted velocity map likely
associated with the red lobe of the main \irsf\ flow.

\subsubsection{Discussion} \label{MGasDisc}
The high-resolution \thco\ data show a wealth of structure never
before seen in this well-studied region. It is clear from this map
that energy from embedded sources has accelerated ambient cloud
gas. Some of this gas resides in the line wing emission. A vast
majority of the features in the optically thin \thco\ line wings
appear as thin arcs of emission. A low
filling factor has been deduced for the L1551 flow by \cite{sne84}
from low-resolution data, and the structure seen in the \thco\ line
wings of Figure~\ref{MGasChmaps} is consistent with their value of
$\sim 0.3$.

The diffusion timescale of distinct features seen in the channel maps
can be estimated by dividing the size of the feature by the
width in velocity space, $t_{\rm diff} \sim \Delta x/\Delta
v$. Typically, $\Delta x \sim 0.05$\,pc (see \S\,\ref{MGasDandR}) and
$\Delta v \sim 1$\,\kms, giving $t_{\rm diff} \sim 10^5$\,years.
The coherence in both position and velocity space of the filamentary
structures seen in the \thco\ line wings along with the short implied
diffusion timescales leads us to interpret them as
the shocked gas boundaries separating quiescent gas from outflow motions
from young sources \citep[see also][]{bar93b}.
The velocity widths of these features do not extend much beyond the
escape velocity of the cloud, $\sim 2$\,\kms, and will likely continue
to decelerate into the ambient gas. The large Reynolds numbers of
these flows imply that they are highly turbulent.

\section{Stellar Feedback} \label{MGasOutflows}
The high-resolution maps of \S\,\ref{sMolGas} show that energy is
being injected into the ambient cloud from where known embedded
sources exist. We look at three particular regions that show clear
signs of stellar feedback---the southwest lobe of the \irsf\ outflow,
the northeast cavity, and the XZ/HL\,Tau region---to gain insight into
the how stellar feedback is affecting the molecular environment of
L1551.

\subsection{The Expanding Shell from the \irsf\ Outflow}
\label{MGasIRS5Outflow} 
\begin{figure}
\centerline{\includegraphics[width=3in]{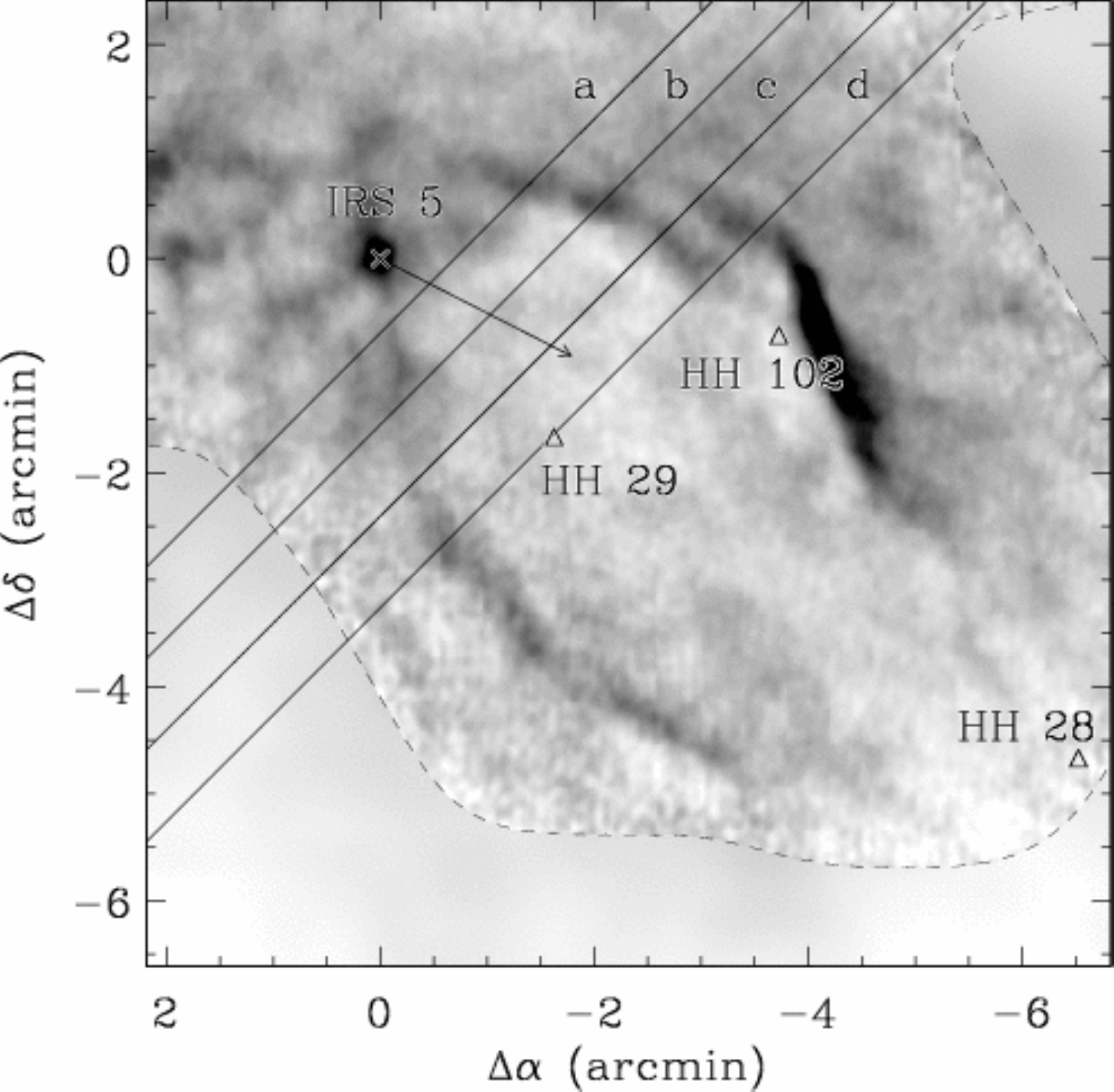}}
\caption{Velocity integrated \thco\ emission as in 
  Figure~\ref{MGasHires13} with the position-velocity cuts through the
  blue lobe of the \irsf\ outflow overlaid ({\it a--d}). The positions
  of \irsf\ and prominent HH objects ({\it triangles}) are labeled,
  and the direction of the \irsf\ jet is shown ({\it
  arrow}). \label{MGasPVCuts}}    
\end{figure}

The blueshifted, or southwest, lobe of the \irsf\ outflow appears as a
thin, limb-brightened shell in Figures~\ref{MGasHires13},
\ref{MGasHires18} and \ref{MGasChmaps}. 
The position-velocity ($p$-$v$) cuts
shown in Figure~\ref{MGasPVCuts} oriented perpendicular to the \irsf\
outflow axis reveal the kinematic nature of this feature. 

\begin{figure}
\centerline{\includegraphics[width=3.4in]{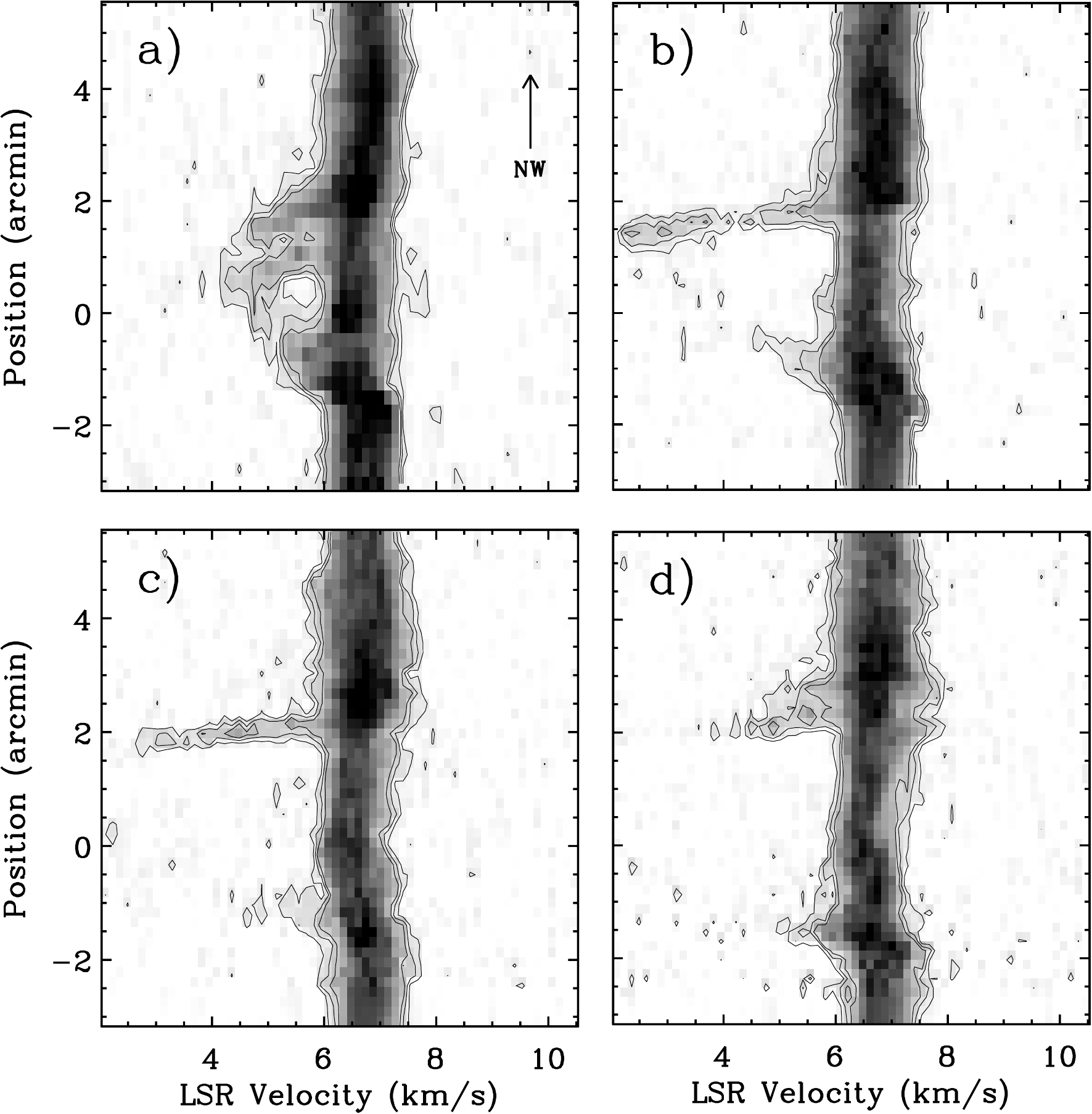}}
\caption[\thco\ emission in the position-velocity cuts across the blue
  lobe of the \irsf\ outflow]{\thco\ emission in the position-velocity
  cuts across the blue 
  lobe of the \irsf\ outflow at a position angle of $135^\circ$,
  roughly perpendicular to the \irsf\ outflow axis. The projected
  distances of the cuts from \irsf\ are $\sim 0^\prime\!\!.5$,
  $1^\prime$, $1^\prime\!\!.5$, and  $2^\prime$ for ({\it a}), ({\it
  b}), ({\it c}), and ({\it d}), respectively (see
  Figure~\ref{MGasPVCuts}). The 
  gray scale follows a linear transfer function from 0.3--7\,K and the
  contour levels are 0.9, 1.5, and 2.1\,K. The zero position roughly
  delineates the outflow axis of symmetry. \label{MGasPVSW}} 
\end{figure}

Figure~\ref{MGasPVSW} shows the emission in $p$-$v$ cuts ({\it
a})--({\it d}) in order of increasing distance from \irsf. Along cut
({\it a}) a full arc of emission blueward of the line core can be
seen. This $p$-$v$ signature is what is expected from an expanding
shell with a velocity of $\sim 2$\,\kms. Moving further away from
\irsf\ in cuts ({\it b}) and ({\it c}), the shell walls appear
spatially thin, $\sim 0.02$\,pc, and have large velocity
widths. Beyond velocities 3--4\,\kms\ from the line core the shell
boundaries become discontinuous in $p$-$v$ space, and in cut ({\it d})
no significant amount of emission appears at high velocity. This suggests
that the expanding shell has burst out the front side of L1551, in
accord with previous interpretations of observational
data~\citep[\eg,][]{dra85,mor88,whi00}. The dynamical age of this
feature from its diameter, $\sim 4^\prime\!.5$, and expansion velocity
$\sim 2$\,\kms, is roughly $10^5$\,years 

The symmetry of this expanding shell is not aligned with then known
direction of the \irsf\ jet~\citep{mun83} as seen in
Figure~\ref{MGasPVCuts}. Yet the shell wall is spatially thin, $\sim
0.02$\,pc, and has a large spread in velocity, $\sim 3$\,\kms,
implying a short dispersion timescale, $\Delta x/\Delta v \sim
10^4$\,years. High velocity \hone\ seen in the outflow lobes of \irsf\
\citep{gio00} may contribute to the acceleration of this shell and its
spatially thin structure.

Summing the emission in the blue lobe at LSR velocities less than
6\,\kms\ gives a total mass of $\sim 3$\,\msun. This is taken to be a
lower limit to the total shell mass since the shell also contains a
component at the line-of-sight velocity of the ambient gas (see
Figure~\ref{MGasPVSW}). The energy in the expanding shell estimated by
converting \thco\ emission to mass (\S\,\ref{sL1551Mass}) and summing
$1/2\,M|v-V_{\rm cen}|^2$ in each velocity channel less than 6\,\kms\
is $\sim 8\times10^{43}$\,ergs.

The total mass traced by molecular emission in the entire blue lobe of
the \irsf\ outflow is higher than the value derived for the
shell~\citep{sto06}. However, the high velocity gas concentrated
toward the center of the outflow lobe seen in CO will most likely end
up in the diffuse interstellar medium rather than the L1551 cloud.
The existence of HH objects beyond the edge of the cloud
\citep[see][]{dev99,mor06,sto06} and the reflection nebulosity
emanating from \irsf\ \citep{dra85} is further evidence that a
fraction of the accelerated ambient gas is being lost from the cloud.

An estimate for the mass of excavated material can be made by
calculating the decrement of \thco\ emission inside the southwest
cavity relative to the ambient cloud. Using Equation~\ref{thcoav}, the
total mass excavated from the southwest lobe is of order $\sim
2$\,\msun. 

\subsection{The Northeast Cavity} \label{MGasNECavity}
The lack of CO isotopologue emission and extinction of background
starlight in the region to the northeast of \lne, the sharp boundary
around this deficit seen in Figures~\ref{MGasHires13} and
\ref{MGasChmaps} \citep[also see][Figure~11]{mor06}, and the presence
of HH\,286 beyond the cloud edge in this
direction~\citep{dev99,mor06,sto06} are all evidence that mass has
been excavated from this region by stellar energetics. This feature
has been seen in previously 
published data~\citep[see][Figure~4; Pound \& Bally 1991, Figure~4; 
Stojimirovi\'{c} et al. 2006, Figure~3]{mor88}, but has never been
interpreted.
\begin{figure}[!b]
\centerline{\includegraphics[width=3.25in,angle=90]{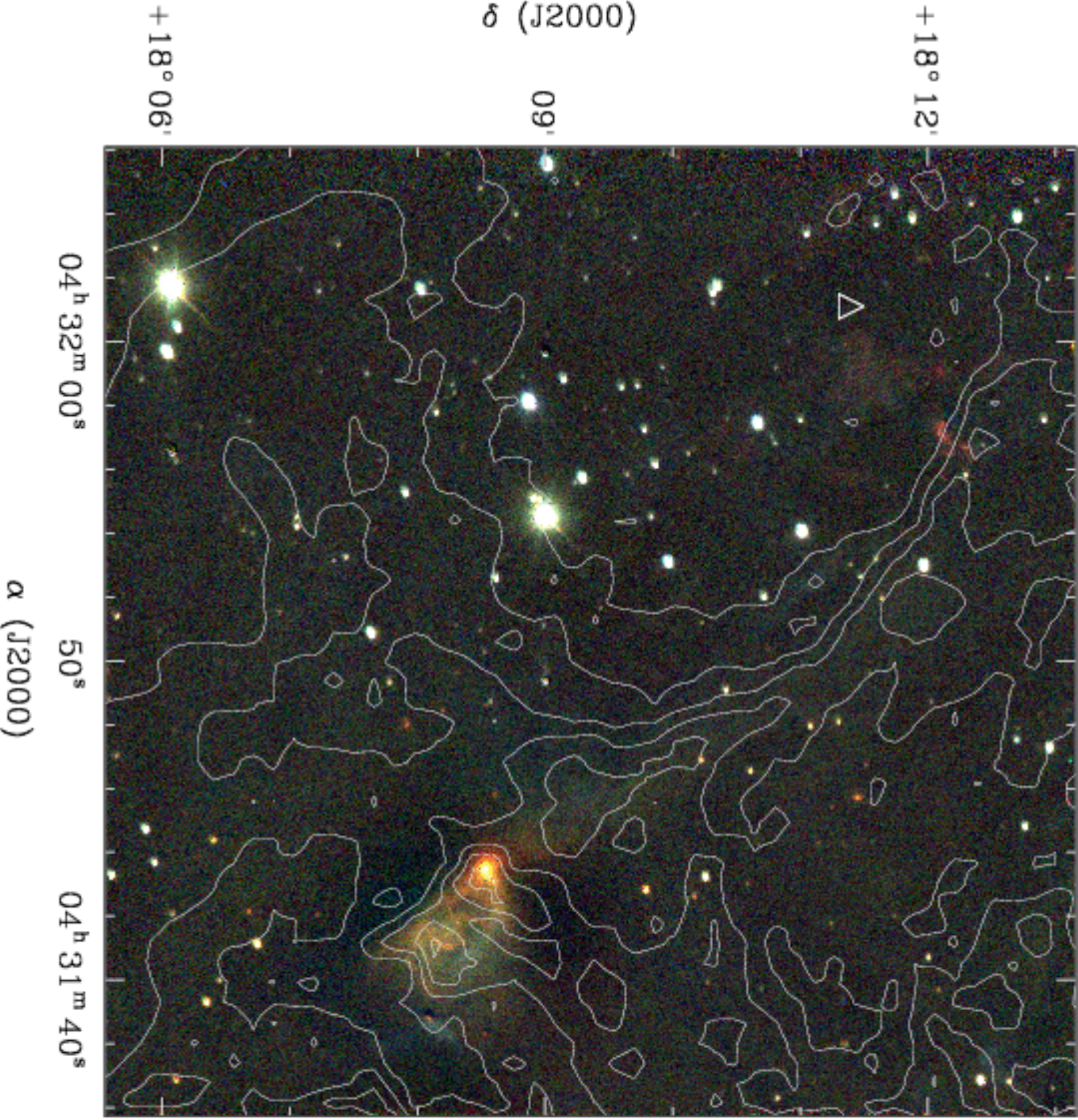}}
\caption{Deep $JHK$ composite image of the \lne\ region overlaid with
  contours of \thco\ emission integrated from 6.5--7.9\,\kms\
  (LSR). The linear stretch on all NIR bands is from $-\sigma$ 
  to $20\sigma$, where $\sigma$ is the rms of the image, and the
  contour levels are $(n+2)$\,K\,\kms, where $n$ is an integer. \lne\
  is the red source to the lower right in
  the image and the position of HH\,262 is labeled ({\it white
  triangle}). \label{MGasLNE}} 
\end{figure}

In Figure~\ref{MGasLNE}, our deep NIR imaging is shown as a $JHK$
composite image overlaid with \thco\ contours. NIR nebulosity on the
southwest side of \lne\ is conical with an opening angle of $\sim
60^\circ$ and is symmetric with the jet axis of $\sim  
243^\circ$ \citep{dev99,reip00}. On the northeast side, the nebulosity
follows the parabolic shape of the \thco\ emission and is brighter
along the northern arm. This nebulosity extends from the \lne\ source
position and has a geometry that implies alignment with the jet axis.

HH\,262 lies within the northeast cavity and has components with a
wide range of radial velocities and proper motions suggesting that the
jets from both \irsf\ and \lne\ may be interacting with this region
\citep{dev99,lop98}. Faint $K$-band emission seen $\sim 1^\prime\!.5$
to the northwest of the nominal HH\,262 position in
Figure~\ref{MGasLNE} is likely due to shocked H$_2$ associated with 
GH\,9~\citep{gra90,lop98}. HH\,286 lies further out of the L1551 cloud
and also originates from either \irsf\ or \lne\ \citep{dev99}. The red
lobe of the \irsf\ molecular outflow overlaps with the northeast
cavity \citep[see][Figure~7]{sto06}, but the axis of that flow lies
along the northern arm of the cavity making \irsf\ unlikely
responsible for excavating the mass from the northeast cavity.

We therefore attribute the northeast cavity feature to \lne\ while,
similar to the blue lobe of the \irsf\ outflow, it contains features
from both \irsf\ and \lne\ \citep{dev99}. The dynamical time for the
northeast cavity based on its projected size and a velocity spread of
$\sim 3$\,\kms\ is $\sim 10^{\rm 4-5}$\,years. The mass of gas
excavated from the region is estimated to be 2--3\,\msun\ from the
deficit of \thco\ and $A_V$ in the cavity relative to the average
column depth of the cloud around the cavity. 

\subsection{The XZ/HL\,Tau Region} \label{MGasXZHL}
The XZ/HL\,Tau region is highly active \citep{mun90}. XZ\,Tau
is known to have poorly collimated outbursts on several year
timescales~\citep{kri99}, and a limb-brightened shell of \thco\ is
centered roughly on its position~\citep{wel00}. In
Figures~\ref{MGasChmaps}({\it a}) and ({\it e}) the acceleration of
the ambient gas from around XZ/HL\,Tau is clearly evident. There is
also a circular region devoid of \thco\ emission in channel map ({\it
a}) centered on the position of LkH$\alpha$\,358 that may be
evidence for a weak wind.  

The expanding half shell seen by \cite{wel00} is evident in channel
maps ({\it a}) and ({\it e}) and extends to the position of
HH\,30\,IRS in map ({\it f}). Also in Figure~\ref{MGasChmaps}({\it f})
there are two arcs curving around the east side of XZ\,Tau that are
disjoint from the lower rim of the XZ\,Tau shell. The near one is
smooth and filamentary, and the far one is clumpy. These features
persist for multiple channels and move in unison away from the
XZ/HL\,Tau region with decreasing velocity. This is consistent with
gas expanding into the ambient cloud as it is accelerated from the
XZ/HL\,Tau region. The expansion velocity is $\sim 0.7$\,\kms\ and the
expansion age is $\lesssim 10^5$\,years assuming this scenario. A
simple estimate for the time between the ejection events is $\sim
10^4$\,years. It is possible that these are molecular signatures of
episodic outbursts similar to those seen in visible
light~\citep{cof04,kri99}. 

\begin{figure}
\centerline{\includegraphics[width=3.25in,angle=90]{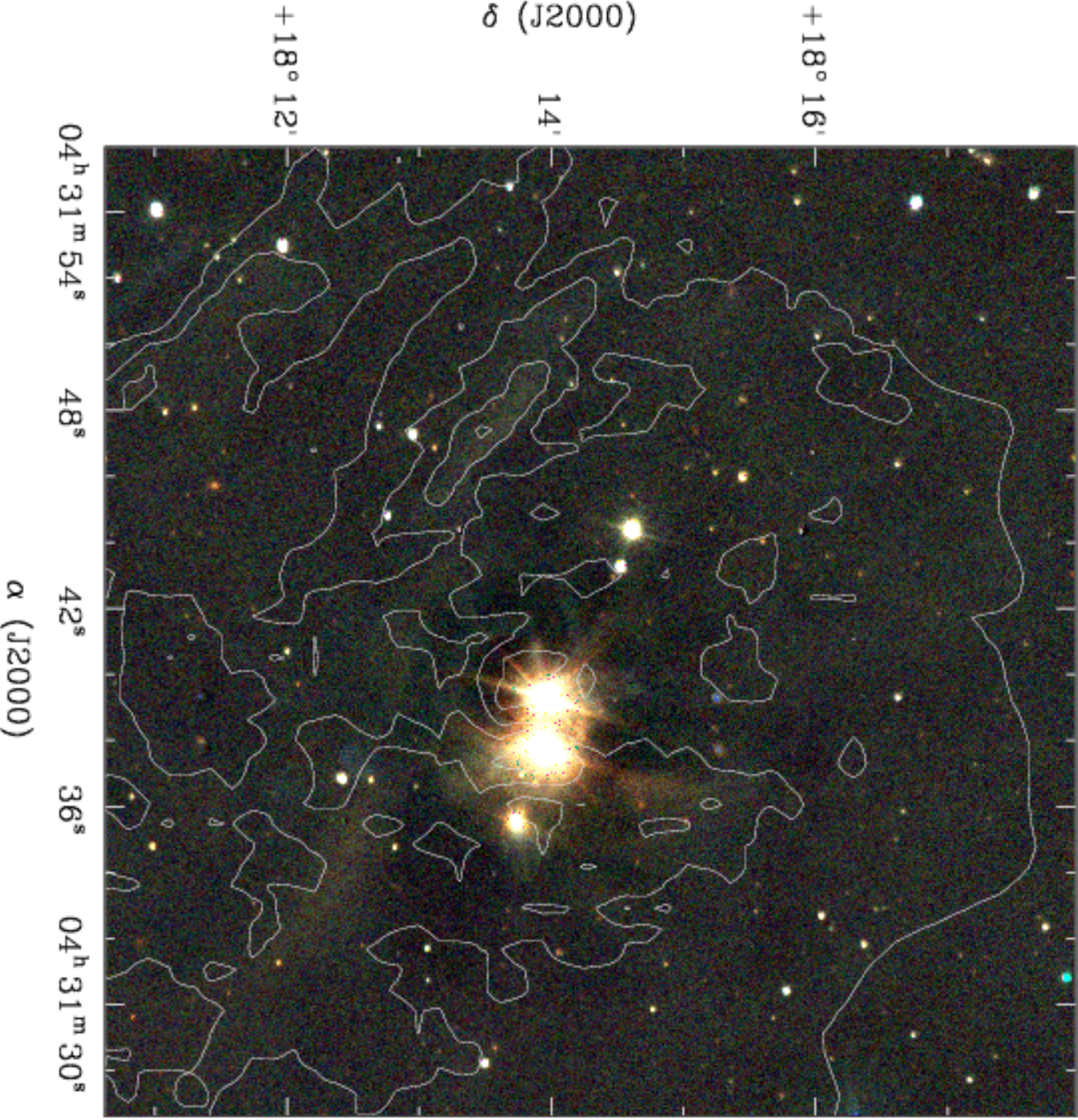}}
\caption{Same as Figure~\ref{MGasLNE}, but for the XZ/HL\,Tau
  region. The \thco\ map has been integrated over velocities between
  5.4 and 7.7\,\kms\ and the contour levels are $(5+ 1.5n)$\,K\,\kms.
  \label{MGasXZHLfig}}    
\end{figure}

On the blueshifted side of the \thco\ line
core in Figures~\ref{MGasChmaps}({\it a}) and ({\it b})
a peculiar linear feature located $\sim 2^\prime$ east of XZ\,Tau
extends in a northeast-southwest direction . This feature persists
over $\sim 2$\,\kms\ and does not vary in position over those
velocities. Figure~\ref{MGasXZHLfig} shows a composite $JHK$ image of
the XZ/HL\,Tau region overlaid with \thco\ contours. This linear
feature seen in \thco\ coincides with NIR nebulosity of the same
morphology that is likely due to reflected light from XZ/HL\,Tau. This
linear feature is also visible in H$\alpha$ emission~\citep{dev99},
and likely delineates the rim of the cavity vacated by the energetics
from XZ\,Tau, HL\,Tau, and HH\,30\,IRS~\citep[see][]{mun90}.  

The \thco\ emission shown in Figure~\ref{MGasXZHLfig} also traces well
other diffuse NIR nebulosity. The nebulosity 
immediately below XZ\,Tau and the nebulosity stretching northward
from HL\,Tau also may be delineating a boundary between high and low
extinction regions in L1551. Unfortunately, not enough background
stars are seen in our deep NIR images to do an extinction analysis in
this region. However, these data are consistent with the
interpretation of \cite{wel00} that the HL\,Tau lies at the boundary
of a region excavated by XZ\,Tau. 

Stars in this region of L1551 must have destroyed part of the cloud
both to expose XZ\,Tau along our line of sight and to create the
features seen in \thco\ and the NIR. However, the modest diminution of
\thco\ column depth seen in our figures translate to only $\sim
0.1$\,\msun\ of excavated matter from the XZ/HL\,Tau region.

\subsection{Summary of Stellar Feedback}
\label{MGasOutflowSum} 
A number of new features are revealed in our high-resolution maps that
show clear signs of interaction between outflow from young
stars and the ambient cloud from which they formed. The \thco\ line
wing emission traces the 
interaction between gas accelerated by embedded sources and the
ambient cloud. At our sensitivity limit, this emission does not extend
to velocities much beyond the escape velocity of the cloud and is
therefore a good tracer of the energy that will remain in the cloud
and ultimately contribute to its turbulent velocity field. 

Using the closely Gaussian \ceo\ composite spectrum to characterize
the quiescent gas, the \thco\ line wings are defined as $V_{\rm red} >
7.5$\,\kms, and $V_{\rm blue} < 5.9$\,\kms\ (LSR). The energy in the
red and blue wings are $\sim 1.5$ and $2.0\times10^{44}$\,ergs,
respectively. If isotropy of outflow motion is assumed, the total
energy in the \thco\ line wings becomes $\sim 10^{45}$\,ergs.

This is about a factor of 5 lower than recent estimates for the
total outflow energy in L1551 using $^{12}$CO \citep{sto06} which
traces both the bound and unbound outflow gas.
All major outflow regions show clear evidence that gas is being
accelerated beyond the gravitational confines of the cloud, and at
least 4--5\,\msun\ of mass has been excavated from L1551 by the
present day embedded stars. This suggests that L1551 was more massive at
the inception of star formation by perhaps 60\%.

\section{Turbulence} \label{MGasTurb}
While our \thco\ data show a wealth of structure in the line wings due
to the energetics from young embedded sources, the \ceo\ emission has
a closely Gaussian profile at our sensitivity level and is optically
thin throughout the map. The \ceo\ emission is thus better suited to
probe the turbulent velocity and density fields in L1551. 

The one-dimensional velocity dispersion of the composite \ceo\
spectrum is 0.31\,\kms. Subtracting in quadrature the thermal
contribution to the measured width using Equation~\ref{sigmatherm}
with $\bar{m} = 30\,m_{\rm H}$ and then multiplying by $\sqrt{3}$, the
three-dimensional turbulent velocity dispersion $\sigma_{\rm turb} = 
0.52$\,\kms. The total turbulent energy in L1551 is estimated to be
$1/2 M_{\rm tot}\sigma_{\rm turb}^2 = 4.3 \times 10^{44}$\,ergs, and
$E_{\rm turb} \approx 0.4\,E_{\rm grav}$. 

\subsection{Spectrum of Fluctuations}
A complete statistical description of fully developed turbulence is
obtained in the scaling relations of the density and velocity 
fields. Under ideal circumstances,
energy is conserved as it cascades from large to small scales at a
rate proportional to $v^3(L)/L$, implying a self-similar velocity
dispersion that scales as $v(L) \propto L^\gamma$, with $\gamma =
1/3$ \citep{kol41}. This line
width-size relation translates to an energy spectrum $E(k) \propto
k^{-\beta}$, with $\beta = 5/3$, and a power spectrum of $P_{\rm v}(k)
\propto k^n$, with $n = {-(\beta + D - 1 )}$ and $D$ being the number
of dimensions that the power spectrum is computed over. Other
turbulence models, such as shock dominated \citep{bur74} and
multi-fractal~\citep{bol02}, also show distinct scaling relations.

Unfortunately, observers can only measure two dimensional
distributions of velocity and density on the plane of the sky, and
many different techniques are used to deduce the three-dimensional
distributions from the available information. Line widths and velocity
centroids are a direct measure of the velocity field and have been
used widely to infer the characteristics of the underlying
turbulence~\citep[\eg,][]{lar81,sca84,dic85,kle85,mie94}. Other
methods including the spectral correlation
function~\citep[SCF;][]{ros99}, principle component
analysis~\citep[PCA;][]{bru02},  
$\Delta$-variance~\citep{stu98}, and velocity channel
analysis~\citep[VCA;][]{laz00,laz04} have had varying degrees of
success.

\subsubsection{Velocity Channel Analysis} \label{VCA}
Density and velocity fluctuations are not independent within the
framework of homogeneous, isotropic turbulence and can theoretically
be discerned through observation. The velocity channel analysis (VCA)
technique has been developed toward this end~\citep{laz00,laz04} and
has been used to successfully separate the velocity and density
contributions to the power spectrum of \hone\ in the Small Magellanic
Cloud \citep{sta01}. 

The intensity fluctuations within a spectral line map integrated over
the full extent of the line profile are dominated by the density
fluctuations such that the index of the two-dimensional power spectrum
of intensity fluctuations equals the index of the three-dimensional
density power spectrum
\begin{equation}
P_{\rho}^{2D, {\rm thick}}(k) \propto P_{\rho}^{3D}(k) \propto k^n
\label{MGasPkthick}
\end{equation}

Intensity fluctuations within a thin slice in velocity space have
contributions from both the velocity and density fluctuations in
proportions that depends on the steepness of the three-dimensional
density spectrum
\begin{equation}
P_{\rho}^{2D, {\rm thin}}(k) \propto \begin{cases} 
  k^{n+\alpha/2}  & : \quad n > -3 \\
  k^{-3+\alpha/2} & : \quad n < -3 \\
\end{cases}
\label{MGasPkthin}
\end{equation}
Here $\alpha$ is the power-law index of the second-order velocity
structure function (see \S\,\ref{sSF}). The formal
criterion for a slice to be thin is that the dispersion of 
turbulent velocities on the scales studied be larger than the velocity
slice thickness. The thin slice regime is not attainable for sub-sonic
turbulence. 

\begin{figure}
\centerline{\includegraphics[angle=90,width=3.4in]{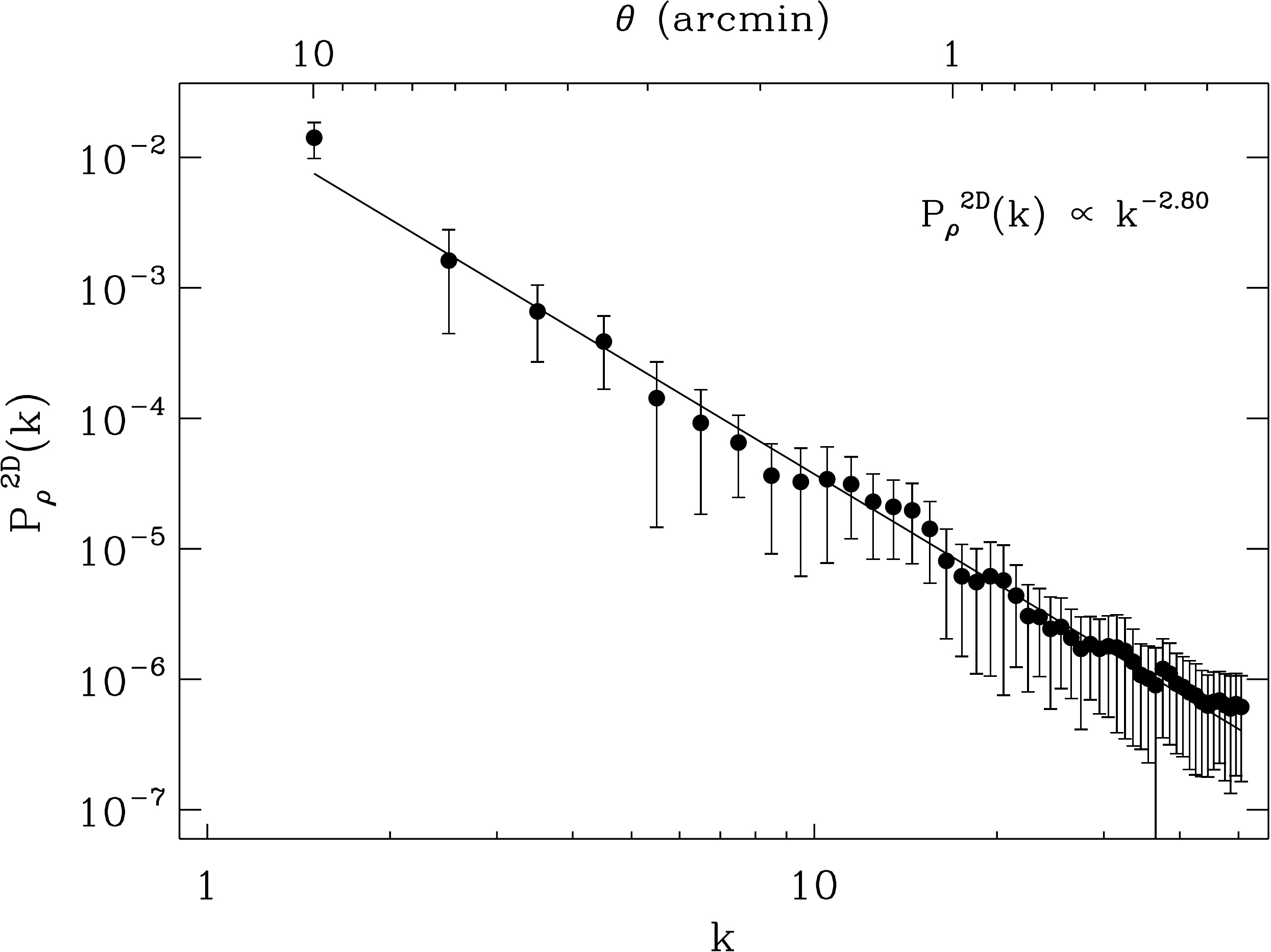}}
\caption[The annular average of the two dimensional density power
  spectrum of \ceo\ emission as a function of wave number]{The annular
  average of the two dimensional density power 
  spectrum of \ceo\ emission as a function of wave number. The
  corresponding scale in the map is plotted on the upper axis. The
  best fit power-law slope for $P_\rho^{\rm 2D}(k)$ is $-2.8$ and is
  shown as the line traversing the data. \label{MGasPS}}
\end{figure}
Figure~\ref{MGasPS} shows the power spectrum of the velocity integrated
\ceo\ within the high-resolution region outlined with the dashed line
in Figure~\ref{MGasHires18}. The velocity integrated map was first
buffered on all sides with zeros to minimize aliasing effects and was
then Fourier transformed using a Fast Fourier Transform (FFT) 
algorithm and squared. Circular annuli are averaged to obtain the mean
value of the power spectrum in each annulus, and the dispersion of
values within each annulus determine the error bars.
\begin{figure}
\centerline{\includegraphics[angle=90,width=3.4in]{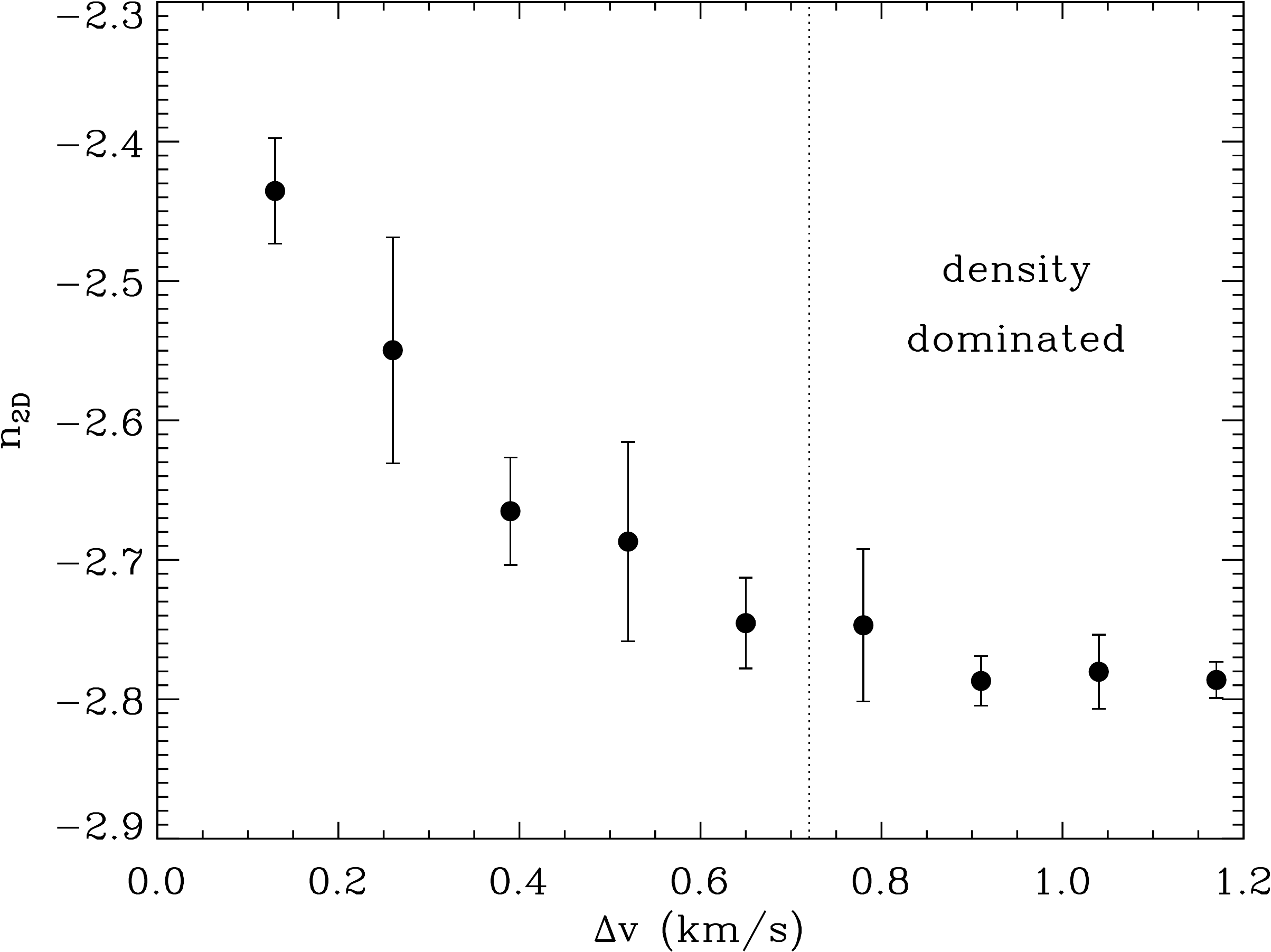}}
\caption{The derived
  power-law index for $P_{\rho}^{2D}(k)$, the 
  two-dimensional power spectrum of density fluctuations, as a
  function of the velocity width of the integrated map. The decrease
  in index as a function 
  of increasing velocity width is expected, and the flattening out at
  large $\Delta v$ signifies the density dominated
  regime. \label{MGasVCA}}  
\end{figure}

The power spectrum of Figure~\ref{MGasPS} follows a power-law form
over the full range of scales sampled. A non-linear least squares fit
to the power spectrum returns a slope of $n = -2.8\pm0.1$,
characterizing the density fluctuations as ``shallow,'' $n > -3$. The
error here is derived from the variance of power spectrum indices fit
to subsets of the \ceo\ data.

Figure~\ref{MGasVCA} shows the slope of the two-dimensional power
spectrum as a function of velocity width. For each width, $\Delta v$,
the \ceo\ map was integrated in velocity across the necessary number of
channels centered around the central channel in the line profile, $V_{\rm
  LSR} = 6.7$\,\kms. A power spectrum was then obtained in the same
manner as described above. Shifting the window of channels by a single
channel width in both directions gives two more measurements of the
slope. The mean of the three values for $n_{\rm 2D}$ is represented
as a filled circle in Figure~\ref{MGasVCA} and the spread
in these three values determines the error bar. 

As the channel width increases, the slope of the two-dimensional power
spectrum monotonically decreases. This is the behavior predicted by
the VCA analysis. The trend levels out for channel widths greater than
$\sim 0.6$\,\kms\ and likely signifies the transition into the density
dominated regime. There is no sign of this 
trend tapering off for small $\Delta v$ and it is possible that this
trend will continue to the thermal width of the \ceo\ line,
0.06\,\kms\ (Lazarian, A. 2006, private communication).

Assuming that the value of $n_{\rm 2D} = -2.44$ is representative of
the slope of $P_{\rho}^{2D}(k)$ in the thin regime, a value for the
velocity spectrum index can be attained using Equation~\ref{MGasPkthin},
$\alpha = 0.72 \pm 0.1$. This implies a three-dimensional power
spectrum of velocity fluctuations, $P_{\rm v}^{3D}(k) \propto
k^{-3.72}$, near the value predicted by the Kolmogorov formalism,
$-11/3$.

\subsubsection{Line Width-Size Relationship} \label{LdV}
\begin{figure}
\centerline{\includegraphics[angle=90,width=3.4in]{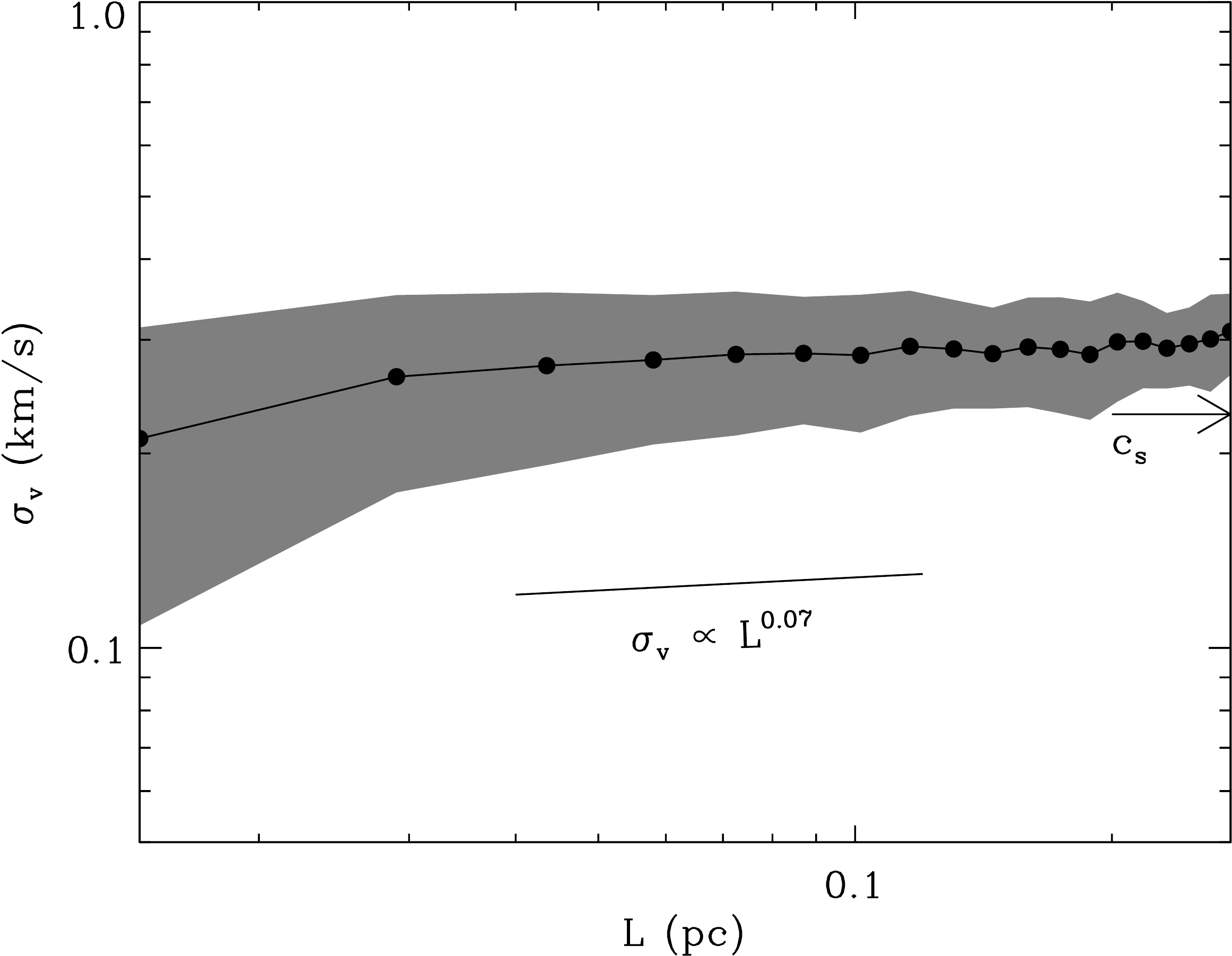}}
\caption{The line width-size relation of the \ceo\ gas in L1551. The
  sound speed for molecular gas with solar helium abundance at a
  kinetic temperature of 15\,K is shown with the
  arrow. \label{MGasLdV}}  
\end{figure}

The VCA results imply an energy spectrum index, $\beta = 1.72$ and a 
line width-size index of $\gamma = 0.36$. The line width size
relationship for our \ceo\ data are shown in Figure~\ref{MGasLdV}. The
one-dimensional line widths in Figure~\ref{MGasLdV} are computed by
taking an intensity weighted 
average of all independent pixels in the \ceo\ map downgraded in resolution to
scale $L$. The average line width is then corrected for cloud depth
effects by subtracting the dispersion in line centroids at the given
map resolution~\citep[see][]{oss02b}. The derived trend in line width
is characterized by $\gamma = 0.07\pm 0.01$. 

This derivation of the line width-size relationship is
fundamentally different than the standard method used where regions
are identified as individual clouds or cores based on prescribed
criteria and their sizes and line widths are then determined for each
separate region~\citep[\eg,][]{lar81,sol87,ful92}. The observed
power-law form of the line width-size relationship is generally
thought to arise from interstellar turbulence~\citep[{\eg},][]{lar81},
but the relationship formed in this manner does not contain any direct
information regarding the nature of turbulence {\it within} individual
clouds.

In L1551, the line width-size relationship is very shallow. This
result is inconsistent with the results from the VCA analysis assuming
that the turbulence is homogeneous and isotropic. However, the derived
line width-size relationship implies straightforwardly that the
observed line widths are originating on small spatial scales which may
indicate the presence of ongoing feedback \citep{nak06}.

\subsubsection{The Structure Function} \label{sSF}
The velocity structure function measures velocity differences as a
function of projected separation and is a common
statistic used to characterize turbulent motions
\citep[{\eg},][]{mie94,bru02,esq05}. For homogeneous, isotropic
turbulence, the structure function is expected to follow a power-law
form, $S_{\rm v}(\vec{\tau}) \propto |\vec{\tau}|^\alpha$. The index,
$\alpha$ is related to the index of the energy spectrum and line
width-size relation, $\alpha = \beta - 1$ and $\alpha = 2\gamma$.

The form of the second-order velocity structure function is
\begin{equation}
S_{\rm v}(\vec{\tau}) = \sum \langle\left[v_{\rm
  c}(\vec{r}) - v_{\rm c}(\vec{r} +
  \vec{\tau})\right]^2\rangle,
\label{Stau}
\end{equation}
where the position, $\vec{r}$, and lag ({\it i.e.}, projected
separation), $\vec{\tau}$, are two-dimensional functions on the plane
of the sky, and the angled 
brackets signify an average over all pairs. The velocity centroid is
\begin{equation}
v_{\rm c}(\vec{r}) = \frac{\sum T(\vec{r},v)v(\vec{r})}{\sum T(\vec{r},v)},
\label{vcen}
\end{equation}
where $T$ is the antenna temperature and the sum is over velocity. The
structure function of Equation~\ref{Stau} normalized by 
the variance of velocity
centroids across the map,
\begin{equation}
\sigma_{\rm c}^2 = \frac{\sum\left[v_{\rm c}(\vec{r}) - \mu\right]^2}{N},
\label{sigmac}
\end{equation}
is denoted  $S_{\rm v}^*(\vec{\tau})$, where $N$ is the number of
pixels in the sum and 
\begin{equation}
\mu = \frac{\sum v_{\rm c}}{N}
\label{meancentroid}
\end{equation}
is the mean velocity centroid of the map. The
normalized structure function has an algebraic relationship to the
spatial autocorrelation function and approaches 
values of 2 for uncorrelated spectra~\citep[see][]{mie94}. 
\begin{figure}[!b]
\centerline{\includegraphics[angle=90,width=3.4in]{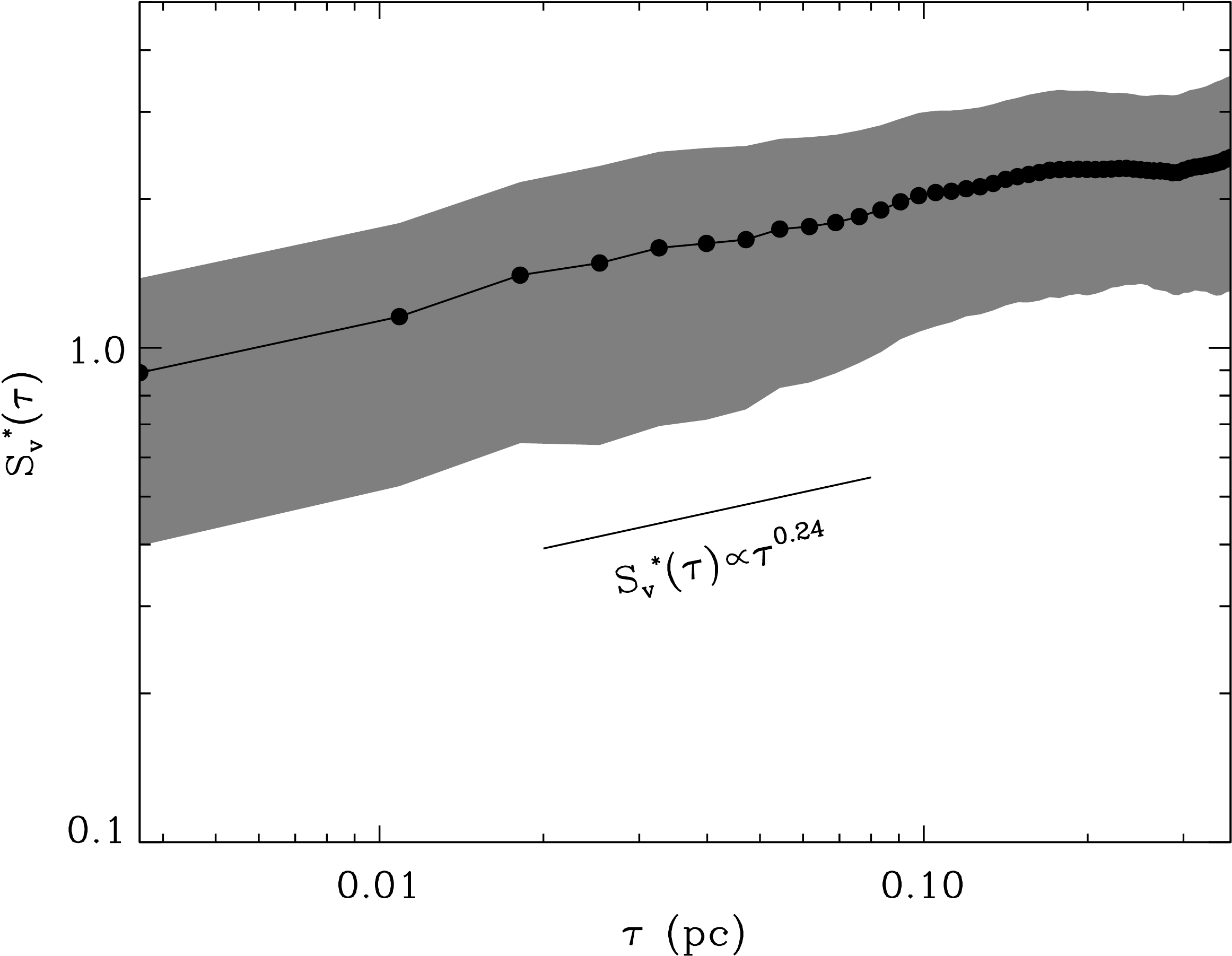}}
\caption{Normalized second-order velocity structure
  function of the 
  \ceo\ emission in L1551 within the high-resolution region outlined
  in Figure~\ref{MGasHires18}. The gray shaded
  region indicates the allowed values of $S_{\rm v}^*(\tau)$ within
  the computed error estimates. The best power-law fit for $\tau <
  0.2$\,pc is shown. \label{MGasSFBIMA}} 
\end{figure}

Detailed numerical tests have shown that the velocity structure function can be
successfully used to retrieve the scaling properties of a
three-dimensional velocity field from a data cube of
position-position-velocity for turbulence with sonic Mach number $M_s
\leq 2.5$ and in regions that are not dominated by density
fluctuations~\citep{bru02,esq05}. A straightforward way to compare the
relative importance of the density fluctuations is through the second
order density structure function.
\begin{equation}
S_\rho(\vec{\tau}) = \sigma_{\rm v}^2\sum\langle\left[I(\vec{r}) - I(\vec{r} +
  \vec{\tau})\right]^2\rangle,
\label{SItau}
\end{equation}
where
\begin{equation}
I(\vec{r}) = \sum T(\vec{r},v).
\label{Idv}
\end{equation}
The sum in Equation~\ref{Idv} is over velocity, and
$\sigma_{\rm v}$ is the velocity dispersion of the gas. To directly
compare Equation~\ref{Stau} and Equation~\ref{SItau}, ``unnormalized''
velocity centroids,
\begin{equation}
\tilde{v}_{\rm c}(\vec{r}) = \sum T(\vec{r},v)v(\vec{r}),
\label{vtildec}
\end{equation}
replace the velocity centroids of Equation~\ref{Stau}. The structure
function of the unnormalized velocity centroids is 
denoted $\tilde{S}_{\rm v}(\tau)$, and it has been determined by numerical
simulation that the velocity structure function accurately traces the
turbulent velocity statistics for $\tilde{S}_{\rm v}(\vec{\tau}) \gg
S_\rho(\vec{\tau})$~\citep{esq05}.

A value for the velocity centroid was computed at each independent
pixel according to Equation~\ref{vcen}. Only channels with signal
greater than 2.5\,$\sigma$ within the full width between first nulls
of the composite spectrum were chosen as part of the sum, and only
spectra with 3 or more channels meeting this criterion were
considered, the rest were masked. The normalized structure function,
$S_{\rm v}^*(\vec{\tau})$, is then computed using $\mu =
6.69$\,\kms\ and $\sigma_{\rm c} = 0.13$\,\kms. A noise 
correction for $\sigma_{\rm c}$ \citep{mie94} has negligible
effect. The dispersion in centroid values at a given lag is
used as the error estimate in $S_{\rm v}^*$ at that lag.

Figure~\ref{MGasSFBIMA} shows $S_{\rm v}^*(\tau)$ computed for \ceo\
over the high resolution region of Figure~\ref{MGasHires18}. There is
a shallow, increasing slope for $\tau \lesssim 0.2$\,pc. 
A power-law fit to $S_{\rm v}^*(\tau)$ in this region yields a slope
of $0.24\pm0.01$. The function $\tilde{S}_{\rm v}(\tau)$,
computed according to Equations~\ref{Stau} and \ref{vtildec} has
amplitudes much greater than $S_\rho(\tau)$
everywhere in our map implying that density fluctuations do not
dominate on any scale. 

\subsubsection{Discussion}
The values of $\alpha$ derived from the VCA analysis, the
line width-size relationship, and the second-order velocity structure
function are, $0.72 \pm 0.1$, $0.14 \pm 0.02$, and $0.24\pm0.01$,
respectively. These values differ significantly according to the
stated errors, but systematic errors may be important.

The VCA analysis is sensitive to gas temperature (Lazarian, A. 2006,
private communication), and therefore varying kinetic temperatures
throughout the cloud could confuse the VCA results. Gas temperatures
are observed to vary between 9 and 25\,K. Additionally, the slope of the
two-dimensional power spectrum for a thin slice derived from
Figure~\ref{MGasVCA} is only a lower limit, since the trend toward
larger slope indices may continue to smaller velocity widths than we
have sampled with our data. However, a larger power spectrum index
would only make the derived $\alpha$ values larger, and hence more
discrepant. 

The distribution of $S_{\rm v}^*(\tau)$ values at each $\tau$ is
skewed due to the full spread in values being of order the mean at
most lags. Using the median instead of the mean steepens the slope
considerably to $\alpha = 0.47 \pm 0.02$. However, this still does not
reconcile the disparate derived values for $\alpha$.  

Lastly, the consistency of these analyses depends upon the turbulence
being homogeneous and isotropic. The structure seen in the \thco\ line
wings in \S\,\ref{MGasOutflows} suggests that the energy input into
the L1551 cloud is neither homogeneous nor isotropic, and it is not
clear if this is a sound assumption for the turbulence in the cloud. 

\subsection{The Dissipation and Replenishment of Turbulent Energy}
\label{MGasDandR}
The total turbulent energy in L1551 is a substantial fraction of the
gravitational energy. However, turbulent energy dissipates on short 
timescales~\citep{sto98,mac04} suggesting that the observed turbulence
in L1551 is not primordial.

Galactic shear is negligible at the galactocentric distance
of L1551 given its size, and no evidence of large scale
rotation is seen in the first moment map of \thco\ emission. There are
no signs of recent supernova 
explosions in the proximity of L1551, nor are there \hii\ regions
nearby. The \thco\ line wings show coherent, thin 
structures that are expected at the early time evolution of
supersonic turbulence~\citep[\eg,][]{por92}. This structure is not
random, but is clearly associated with the young stars in
the region.

The total energy in the \thco\ line wings estimated in
\S\,\ref{MGasOutflowSum} is $4$--$10\times10^{44}$\,ergs. Both the
dynamical timescale (\S\,\ref{MGasOutflows}) and 
and the diffusion timescale (\S\,\ref{MGasDisc}) for the
line wing features are $\sim 10^5$\,years, and the
age of the \irsf\ outflow has been estimated to be the same order of 
magnitude~\citep{mor88,ric00}. Therefore the observed energy in the
cloud has been injected over the past $\sim 10^5$\,years,
and the observed rate of energy input into L1551 is $\dot{E}_{\rm
input} \approx 2\times10^{32}$\,ergs\,s$^{-1}$, or 0.05\,\lsun.
\begin{figure}[!b]
\centerline{\includegraphics[angle=90,width=3.4in]{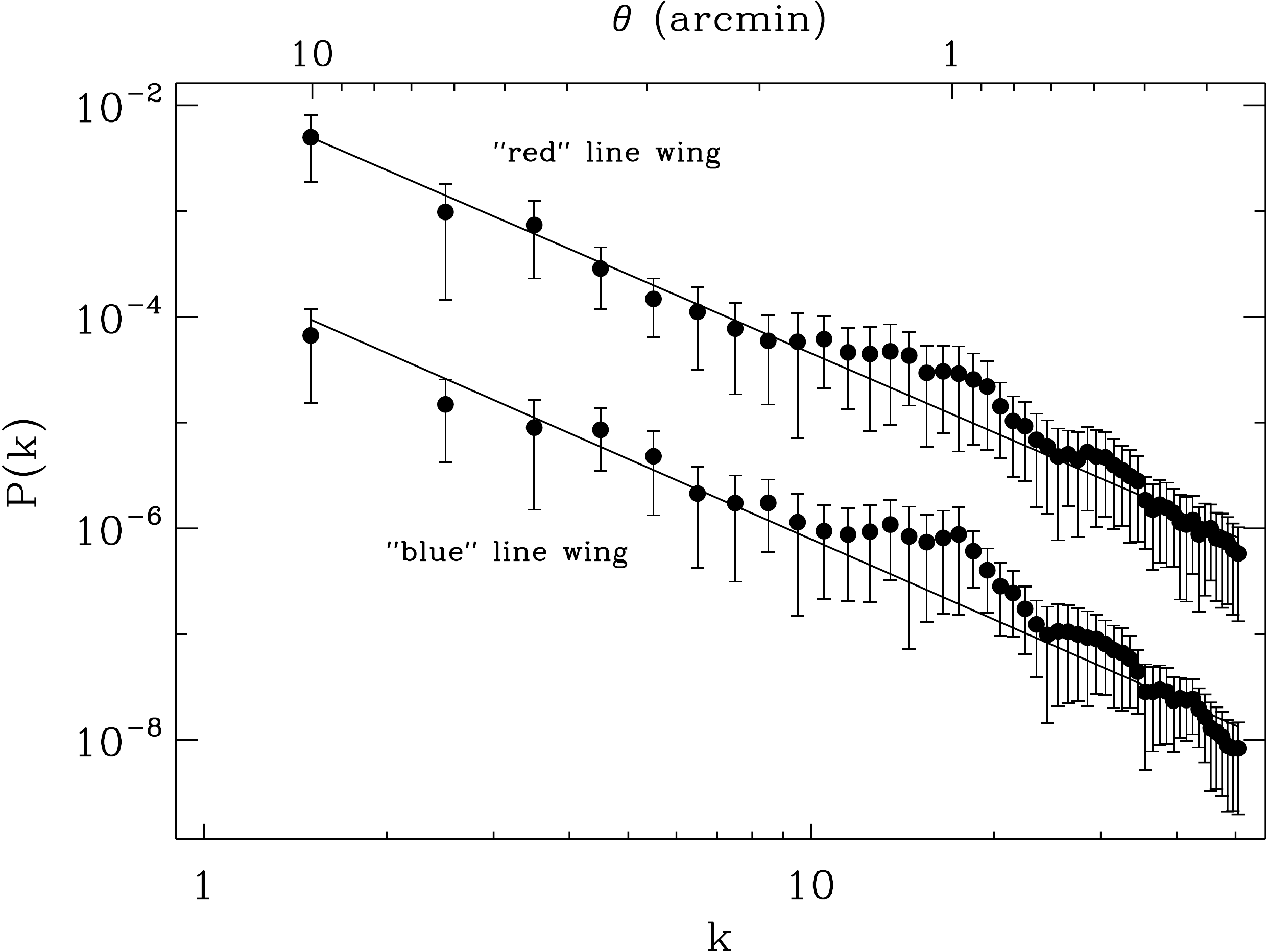}}
\caption{Intensity power spectra for the integrated intensity in the
  red ({\it upper}) and blue ({\it lower}) line wings of the \thco\
  emission. The power spectrum of the blue line wing is divided by 100
  for viewing purposes. The velocity intervals for the integrated
  intensity are 7.5--9.0\,\kms\ and 3.4--5.7\,\kms\ for the red and
  blue line wing data, respectively.  Clear excursions from the
  overall power-law form ({\it lines}) are seen near 1\arcmin\ scales.  
  \label{MGasPSredblue}} 
\end{figure}

The energy from stellar feedback is expected to be highly turbulent
and therefore the turbulent dissipation timescale, $t_{\rm diss} =
E_{\rm turb}/\dot{E}_{\rm diss}$,
sets the rate at which this energy is lost from the cloud. The
dissipation of turbulent energy depends on the scale at which that
turbulence is
driven~\citep{mac99,sto98,gam96}. Figure~\ref{MGasPSredblue} 
shows the power spectra of the integrated \thco\ intensity in the red
and blue line wings. The overall best fit for a single power law is
shown as the thin lines. The slopes of these spectra are more shallow
than the slopes in the line core, and there are also clear deviations
from the linear trend at scales $\sim 1$\arcmin. Since the structures
that are clearly associated with embedded sources dominate the
emission in the line wings, the broad excess peaks in the line wing
power spectra signify a preferential scale at which energy is being
injected into the cloud.  

Using the dispersion of the composite \ceo\ spectrum, $0.31$\,\kms, we
estimate the flow crossing time at an energy injection scale of
$0.05$\,pc, $t_{\rm f}(\lambda_{\rm peak}) \approx 0.1$\,Myr. This
timescale sets the overall scaling for $t_{\rm diss}$. In the
numerical studies by \cite{sto98}, $t_{\rm diss}$ was found to vary
between $\sim 0.5$--0.8\,$t_{\rm f}$ for a range in magnetic field
strength between 0 and $\sim 50$\,$\mu$G using a kinetic temperature
of 15\,K and a particle density equal to $\langle n \rangle \approx
10^3$\,\cmt. For $t_{\rm diss} = 0.7\,t_{\rm f}$ and $E_{\rm turb} = 
4.3\times10^{44}$\,ergs, the turbulent energy dissipation rate in
L1551 is $\dot{E}_{\rm diss} \approx 2 \times
10^{32}$\,ergs\,s$^{-1}$, approximately equal to the estimated energy
input rate from outflow. Using Equation~7 from \cite{mac99} with the
appropriate input values gives a consistent number for
$\dot{E}_{\rm diss}$. 

The uncertainty in the stellar feedback energy is roughly a factor of
2 and the timescale over which the observed energy in outflow has been
injected into the cloud is accurate to roughly the same level. The
turbulent energy figure is relatively accurate, but our
derived dissipation timescale may be off by a factor of 2 or more
based on the broadness of the peaks seen in Figure~\ref{MGasPSredblue}
and the uncertain magnetic field strength in L1551 (see
\S\,\ref{sDisc}). 
Therefore we can only conclude that $\dot{E}_{\rm diss} \approx 
\dot{E}_{\rm input}$ to order of magnitude. However, this rough
agreement between energy rates along with the lack of evidence for any
other form of energy input, the age of L1551 being several dynamical
times, and the fact that turbulent energy only survives for roughly a
flow crossing time, strongly suggest that the observed turbulent
energy has been supplied by outflows from forming stars.

\section{Discussion}\label{sDisc}
The evolution of the L1551 cloud is influenced significantly by the
stars forming within it. Ample evidence for energy injection and cloud
destruction is apparent in our high-resolution maps. The observed
turbulent motions in the cloud are short lived, $t_{\rm diss}
\lesssim 0.1$\,Myr, and exist primarily on small scales where they may
act as a turbulent pressure to support the cloud against gravity
\citep{bon92}. Additionally, the energy input rate is estimated to be
of the order of the dissipation rate. These calculations all support
the idea that the progression of star formation is significantly
influenced by the forming stars \citep[{\eg}][]{nor80,li06}. 

The observed star formation efficiency, SFE, is defined as the present
day stellar mass divided by the total present day mass in stars and
gas. In L1551, ${\rm SFE} \approx 12$\%. The total star formation
efficiency, SFE$_{\rm tot}$, is the fraction of total initial cloud
mass that will ever become stars and may be as low as 9\% in L1551 if
the initial mass of the cloud were $\sim 1.6$ times the mass of the
present day cloud. It appears that L1551 is not capable of producing
significantly more stars in the future limiting SFE$_{\rm tot}
\lesssim 15$\%. However, further study is needed in the high column
regions to the northwest. The theoretical expectation of SFE$_{\rm
tot}$ from outflow-regulated star formation is in the range 25--75\% 
\citep{mat00}. 

The age spread of PMS stars, fractional numbers of protostars, wTTSs,
and cTTSs, and the spatial and kinematic distributions of the stellar
component suggest L1551 has been forming stars for much longer
than its present day dynamical time. The near equality of stellar
feedback and turbulent dissipation provides a means for the cloud to
stay close to virial equilibrium over its stellar production history
\citep[{\eg},][]{tan06,kru06a}. However, the rate of 
star formation has not remained constant in L1551. Roughly 9
protostars have formed in L1551 in the past 
$\sim 0.1$\,Myr. If this rate of star formation were projected
backward in time even a few million years, the number of stars
produced in L1551 would be in the hundreds. Therefore, we conclude
that the increase in the rate of 
star formation over the lifetime of L1551 is robust. Evidence
for accelerating star formation has been seen in Taurus as a
whole~\citep{pal02} and in other well-studied star forming
regions~\citep{pal00,huf06} and is expected if the star formation rate
follows the dynamical rate of a contacting cloud~\citep{pal99}.

Accelerating star formation fits loosely within the framework of
dynamical star formation \citep[{\eg},][]{elm00}. However,
dynamical models typically neglect or discount the
effects of stellar feedback. This is untenable in the case of
L1551. The timescales of purely dynamical star formation, typically
1--$2\,t_{\rm dyn}$, also do not match our observations. Even in the
likely case that the dynamical timescale of the initial cloud was
longer than the present day estimation, clear evidence for the
formation of another star or stellar system that has yet to form
pushes the productivity lifetime of L1551 well beyond the limits of
purely dynamical formation. 

Missing from this picture of star formation is information regarding
the magnetic field. Polarization measurements of background starlight
in the periphery of L1551 reveal a magnetic field direction somewhat
suggestively aligned with the \irsf\ outflow direction. However, a
sub-millimeter polarization measurement of the dusty clump around
\irsf\ suggests a field direction nearly perpendicular to the outflow
axis \citep{val00}. And while a firm lower limit of $2\,\mu$G has been
set for the magnetic field strength in L1551 via rotation measures of
background extragalactic radio sources \citep{sim86}, useful
constraints on the field strength are precluded without detailed
information about the ionization state of the gas. 

The relative importance of the magnetic field in determining the
structure and evolution of L1551 is presently not clear. More detailed
observations regarding the magnetic field strength and orientation in
L1551 are necessary to compliment this and other work
\citep[{\eg},][]{mor06,sto06} and to assess the role of
magnetic fields in the formation of
stars~\citep[{\eg},][]{shu87,mck89,cio93,shu04}. 

\section{Conclusion} \label{sConc}
This article outlines an extensive observational case study of an
active star-forming molecular cloud and its pre-main sequence
association of stars. New, key observational additions to the study of
this region include wide-field, high-resolution maps of \thco$(1-0)$
and \ceo$(1-0)$ created through the combination of single-dish and
interferometric data; deep, wide-field, near-infrared
imaging; Spitzer IRAC data; and the culling of 2MASS and literature
data regarding the young stellar population in the region. These data
shed light on the relationship between the diverse phenomena observed
in L1551 and the progression of star formation in this cloud.

Within a $1.\!\!^\circ6$ radius of the \ceo\ intensity
weighted mean position of L1551, $4^{\rm h}31^{\rm m}24^{\rm s}$,
$+18^\circ10^\prime00^{\prime\prime}\,\,(J2000)$, 35 PMS
stars are selected to be part of the L1551 association. The
distribution of spectral class and derived bolometric luminosities
suggest that star formation began $\gtrsim 6$\,Myr ago. The spatial
and kinematic distribution of PMS stars are also consistent with a
long star formation history. The relatively high current rate of star
formation is clear evidence that the star formation rate has increased
over the productivity lifetime of L1551, consistent with accelerated
star formation scenarios. The star formation appears to be progressing
east to west with individual sites of star formation likely occurring
in regions of quiescent gas. 

The L1551 cloud contains 160\,\msun\ of total mass as traced by dust
extinction, factors of 1.4 to 2 higher than previous estimates from
molecular emission. The
discrepancies are likely due to the saturation of CO emission in high
column regions and the
destruction of CO in areas of low extinction. The dynamical timescale
for the cloud, given the  radial distribution of mass within its
0.9\,pc radius, is $t_{\rm dyn} = 1.1$\,Myr. This is significantly
shorter than the inferred age of L1551 based on its stellar component.

The composite \ceo\ line width infers a total kinetic energy in the
cloud in accord with virial equilibrium assumptions. The turbulent
velocity field in L1551 produces a density power spectrum $P_{\rho}(k)
\propto k^{-2.8}$. We derive an energy spectrum index $\beta \approx
1$--2, but the observed turbulent motions exist primarily on small
scales as inferred directly from the line width-size relationship in
the cloud. The decay timescale for this turbulent energy is short,
$t_{\rm diss} \lesssim 0.1$\,Myr and is not primordial in origin.

The high-resolution \thco\ emission traces well the stellar
feedback in the cloud. Thin, filamentary features in the line wings of
the \thco\ emission signify the interface between outflow and
quiescent gas and are used to estimate the stellar feedback,
$\dot{E}_{\rm input} \approx 0.05$\,\lsun. This is not the full
mechanical luminosity of outflow in L1551, rather a significant
fraction of outflow energy contributes to the destruction of the
cloud or is otherwise lost to the greater interstellar medium. It is
estimated that 4--5\,\msun\ of gas has been excavated 
from the cloud by the current generation of outflows. The preferred
scale at which energy is injected into the cloud is $\lambda_{\rm
peak} \approx 0.05$\,pc, about a thirtieth of the diameter of the
cloud as seen in dust extinction. 

The dissipation rate of turbulence and the rate of energy injection
from outflow balance according to our calculations. There are no other
significant sources of energy seen in L1551, and therefore outflows
provide a significant supply of kinetic energy and play an important
role in the evolution of the cloud. 

\acknowledgments
This work is based in part on observations made with the Spitzer Space
Telescope, which is operated by the Jet Propulsion Laboratory,
California Institute of Technology under a contract with NASA. Support
for this work was provided by NASA through an award issued by
JPL/Caltech. This publication makes use of data products from the Two
Micron All Sky Survey, which is a joint project of the University of 
Massachusetts and the Infrared Processing and Analysis
Center/California Institute of Technology, funded by the National
Aeronautics and Space Administration and the National Science
Foundation. Thanks to Anthony Gonazles for providing the beautiful
FLAMINGOS data. J. S. benefited greatly from numerous discussions
with Steve Stahler, Chris McKee and Alex Lazarian regarding this
research.

Facilities: \facility{BIMA, Spitzer, Kitt Peak 2.1\,m, 2MASS}
\bibliographystyle{/Users/js/Astronomy/references/apj}
\addcontentsline{toc}{chapter}{Bibliography} 
\bibliography{/Users/js/Astronomy/references/references}

\LongTables

\newpage

\centering

\renewcommand{\arraystretch}{1.25}

\begin{deluxetable}{lccccccccccc}
\tablecaption{Pre-main Sequence Stars in L1551 Region}
\tablehead{ \colhead{Name} &
\colhead{$\alpha(J2000)$\tablenotemark{a}} &
\colhead{$\delta(J2000)$\tablenotemark{a}} &
\colhead{$d$\tablenotemark{b} (arcmin)} &
\colhead{Sp. Typ.} &
\colhead{$J$\tablenotemark{a}} &
\colhead{$H$\tablenotemark{a}} &
\colhead{$K_s$\tablenotemark{a}} &
\colhead{$\Delta\alpha$\tablenotemark{c}} &
\colhead{$\Delta\delta$\tablenotemark{c}} &
\colhead{R.V.\tablenotemark{d}} &
\colhead{note\tablenotemark{e}} }
\startdata
  L1551~IRS~5 (A+B)\dotfill & 4:31:34.1 &  18:08:04.9 &   3.08 &   G-K\tablenotemark{1} &   13.71 &   11.51 &    9.82 & \nodata & \nodata & \nodata & EB \\
  HH30~IRS\dotfill &   4:31:37.5 &  18:12:24.5 &   4.01 &    M0\tablenotemark{2} &   15.18 &   14.24 &   13.34 & \nodata & \nodata & \nodata &       EC  \\
  LkH$\alpha$~358\dotfill &    4:31:36.1 &  18:13:43.3 &   4.70 &  M5.5\tablenotemark{1} &   12.79 &   10.92 &    9.69 & \nodata & \nodata & \nodata &       E  \\
  L1551~NE (A+B)\dotfill & 4:31:44.4 & 18:08:31.5 & 5.07 &                \nodata &   16.61 &   13.63 &   11.41 & \nodata & \nodata & \nodata & EB \\
  HL~Tau\dotfill &      4:31:38.4 &  18:13:57.7 &   5.23 &    K7\tablenotemark{1} &   10.62 &    9.17 &    7.41 &    -0.3 &    -2.1 & \nodata &       EC  \\
  XZ~Tau~A\dotfill &     4:31:40.1 &  18:13:57.2 &   5.50 &    M3\tablenotemark{3} &    9.39 &    8.15 &    7.29 &     1.1 &    -1.9 & \nodata &       EB \\
  XZ~Tau~B\dotfill &     4:31:40.1 &  18:13:57.2 &   5.50 &  M1.5\tablenotemark{3} & \nodata & \nodata & \nodata & \nodata & \nodata & \nodata &       EB \\
  MHO~4\dotfill &       4:31:24.1 &  18:00:21.5 &   9.64 &    M6\tablenotemark{5} &   11.65 &   10.92 &   10.57 & \nodata & \nodata & \nodata & \nodata  \\
  MHO~9\dotfill &       4:31:15.8 &  18:20:07.2 &  10.31 &    M4\tablenotemark{5} &   11.21 &   10.55 &   10.30 & \nodata & \nodata & \nodata & \nodata  \\
  MHO~5\dotfill &       4:32:16.1 &  18:12:46.4 &  12.68 &    M6\tablenotemark{5} &   11.07 &   10.39 &   10.06 & \nodata & \nodata & \nodata & \nodata  \\
  V~710~Tau~B\dotfill &   4:31:57.8 &  18:21:35.1 &  14.09 &    M3\tablenotemark{1} &   10.20\tablenotemark{1} &    9.23\tablenotemark{1} &    8.82\tablenotemark{1} &  \nodata & \nodata & 6.73 &       BC  \\
  V~710~Tau~A\dotfill &   4:31:57.8 &  18:21:38.1 &  14.13 &    M1\tablenotemark{1} &    9.82\tablenotemark{1} &    9.06\tablenotemark{1} &    8.69\tablenotemark{1} &    1.5 &   -0.6 &    7.33 &       BC  \\
  V~826~Tau\dotfill &   4:32:15.8 &  18:01:38.7 &  14.88 &     K7\tablenotemark{1} &    9.07 &    8.43 &    8.25 &    1.3 &   -2.2 &    4.43 &      B  \\
  V~827~Tau\dotfill &   4:32:14.6 &  18:20:14.7 &  15.79 &     K7\tablenotemark{1} &    9.17 &    8.49 &    8.23 &    0.8 &   -1.5 &    4.63 & \nodata  \\
  L1551-51\dotfill &   4:32:09.3 &  17:57:22.8 &  16.59 &    K7\tablenotemark{1} &    9.70 &    9.06 &    8.85 &    1.1 &   -1.8 &    5.43 & \nodata  \\
  TAP~44\tablenotemark{6}\dotfill &   4:32:37.7 &  18:06:36.6  &  17.83 & \nodata &   14.22 &   13.83 &   13.72 & \nodata & \nodata & \nodata & \nodata  \\
  RX~J0432.6+1809\dotfill & 4:32:41.1 & 18:09:23.9 & 18.32 &  M5\tablenotemark{4} &   11.31 &   10.65 &   10.34 & \nodata & \nodata & \nodata & \nodata  \\
  UX~Tau~A\dotfill &   4:30:04.0 &  18:13:49.4 &  19.38 &      K5\tablenotemark{3} &    8.62 &    7.96 &    7.55 &     0.7 &    -1.4 &    2.23 &       B  \\
  UX~Tau~B\dotfill &   4:30:03.6 &  18:13:49.5 &  19.47 &      M2\tablenotemark{3} &    9.87 &    8.95 &    8.92 & \nodata & \nodata &    4.63 &       BC  \\
  UX~Tau~C\dotfill &   4:30:03.6 &  18:13:49.5 &  19.47 &      M5\tablenotemark{3} & \nodata & \nodata & \nodata & \nodata & \nodata &    1.33\tablenotemark{12} &       B  \\
  L1551-55\dotfill &   4:32:43.7 &  18:02:56.3 &  20.21 &    K7\tablenotemark{1} &   10.16 &    9.46 &    9.31 &    1.6 &   -1.3 &    4.73 & \nodata  \\
  MHO~6\dotfill &   4:32:22.1 &  18:27:42.6 &  22.45 &        M5\tablenotemark{5} &   11.71 &   11.02 &   10.65 & \nodata & \nodata & \nodata & \nodata  \\
  MHO~7\dotfill &   4:32:26.3 &  18:27:52.1 &  23.19 &        M5\tablenotemark{5} &   11.18 &   10.37 &   10.17 & \nodata & \nodata & \nodata & \nodata  \\
  TAP~47\tablenotemark{6}\dotfill &   4:33:14.1 &  18:16:09.2 &   26.86 & \nodata &    9.62 &    9.11 &    8.95 & \nodata & \nodata & \nodata & \nodata  \\
  DM~Tau\dotfill &   4:33:48.7 &  18:10:10.0 &  34.37 &       M1\tablenotemark{1} &   10.44 &    9.76 &    9.52 &     1.1 &    -1.9 &    3.53 & \nodata  \\
  RX~J0433.7+1823\dotfill & 4:33:42.0 & 18:24:27.4 & 35.81 &  G6\tablenotemark{4} &    9.84 &    9.42 &    9.27 &    -1.3 &    -0.6 & \nodata &       M \\
  HN~Tau~A\dotfill &   4:33:39.4 &  17:51:52.4 &  36.94 &      K5\tablenotemark{3} &   10.82\tablenotemark{1} &    9.49\tablenotemark{1} &    8.44\tablenotemark{1} &     1.5 &    -1.4 & \nodata & \nodata  \\
  HN~Tau~B\dotfill &   4:33:39.4 &  17:51:52.4 &  36.94 &      M4\tablenotemark{3} &   12.62\tablenotemark{1} & 12.05\tablenotemark{1} & 11.62\tablenotemark{1} & \nodata & \nodata & \nodata & \nodata  \\
  TAP~48\tablenotemark{6}\dotfill &   4:33:28.1 &  17:47:06.6 &  37.35 & \nodata &   12.69 &   12.36 &   12.24 & \nodata & \nodata & \nodata & \nodata  \\
  RX~J0432.8+1735\dotfill &   4:32:53.2 &  17:35:33.8 &  40.45 &     M2\tablenotemark{4}  &   10.00 &    9.23 &    9.02 & 0.8 & -1.6 & \nodata & \nodata  \\
  GG~Tau~Aa\dotfill & 4:32:30.4 &  17:31:40.6 &  41.45 &     K7\tablenotemark{7} &    9.24\tablenotemark{7} &    8.27\tablenotemark{7} &    7.73\tablenotemark{7} &     1.4 &    -2.0 &    4.23 & B  \\
  GG~Tau~Ab\dotfill & 4:32:30.4 &  17:31:40.6 &  41.45 &   M0.5\tablenotemark{7} &   10.12\tablenotemark{7} &    9.07\tablenotemark{7} &    8.53\tablenotemark{7} & \nodata & \nodata & \nodata & B  \\
  GG~Tau~Ba\dotfill & 4:32:30.3 &  17:31:30.3 &  41.60 &     M5\tablenotemark{7} &   11.48\tablenotemark{7} &   10.63\tablenotemark{7} &   10.20\tablenotemark{7} & \nodata & \nodata & \nodata & B  \\
  GG~Tau~Bb\dotfill & 4:32:30.3 &  17:31:30.3 &  41.60 &     M7\tablenotemark{7} &   13.16\tablenotemark{7} &   12.38\tablenotemark{7} &   12.01\tablenotemark{7} & \nodata & \nodata & \nodata & B \\
  HBC~407\dotfill &   4:34:18.0 &  18:30:06.7 &  45.93 &     G8\tablenotemark{8} &   10.58 &   10.08 &    9.90 & 0.0 & -0.7 &    5.03 & B \\
  J2-2041\dotfill &   4:33:55.5 &  18:38:39.1 &  45.96 &   M3.5\tablenotemark{9} &   10.53 &    9.87 &    9.61 &     1.0\tablenotemark{9} &    -1.7\tablenotemark{9} & \nodata & \nodata  \\
  RX~J0432.7+1853\dotfill & 4:32:42.4 & 18:55:10.2 & 48.84 & K1\tablenotemark{11} &    9.24 &    8.76 &    8.69 &    -0.5 &    -1.2 & \nodata & \nodata \\
  TAP~51\dotfill &   4:35:14.2 &  18:21:35.6 &  55.87 &      F8\tablenotemark{13} &    9.32 &    9.00 &    8.92 &     0.0 &    -1.2 & \nodata & \nodata  \\
  HBC~412\dotfill &   4:35:24.5 &  17:51:43.0 &  60.03 &     M2\tablenotemark{8} &   10.03 &    9.33 &    9.08 & \nodata & \nodata &    7.33 & B  \\
  HBC~388\dotfill &   4:27:10.6 &  17:50:42.6 &  63.26 &     K1\tablenotemark{8} &    8.78 &    8.39 &    8.30 &     0.2 &    -1.5 &    1.63 & B \\
  HBC~392\dotfill &   4:31:27.2 &  17:06:24.9 &  63.59 &     K5\tablenotemark{8} &   10.28 &    9.71 &    9.50 & 1.0 & -2.8 &    3.13 & B  \\
  RX~J0433.5+1916\dotfill & 4:33:34.7 & 19:16:48.9 & 73.63 & G6\tablenotemark{11} &   10.96 &   10.52 &   10.40 &     0.3 &    -0.7 & \nodata & \nodata  \\
  TAP~37\tablenotemark{6}\dotfill &   4:30:21.5 &  16:55:49.9 &  75.65 &  \nodata &   14.38 &   14.19 &   13.91 & \nodata & \nodata & \nodata & \nodata  \\
  RX~J0437.5+1851\dotfill &  4:37:26.8 &  18:51:22.5 &  95.44 &           \nodata &    9.97 &    8.76 &    9.07 & \nodata & \nodata & \nodata & \nodata  \\
  RX~J0437.4+1851A\dotfill & 4:37:26.9 & 18:51:26.8 & 95.49 & K6\tablenotemark{4} &    9.42 &    8.56 &    8.67 & \nodata & \nodata & \nodata &      BC  \\
  RX~J0437.4+1851B\dotfill & 4:37:26.9 & 18:51:26.8 & 95.49 & M0.5\tablenotemark{4} & \nodata & \nodata & \nodata & \nodata & \nodata & \nodata & B  \\
  TAP~33\tablenotemark{6}\dotfill &   4:22:53.0 &  17:34:15.8 & 126.73 & \nodata &   10.38 &    9.71 &    9.49 & \nodata & \nodata & \nodata &       S  \\
  TAP~53\tablenotemark{6}\dotfill &   4:36:10.8 &  16:04:52.5 & 142.66 & \nodata &   10.43 &   10.01 &    9.98 & \nodata & \nodata & \nodata &       S  \\
  RX~J0431.4+2035\dotfill & 4:31:30.4 & 20:35:38.4 & 145.65 & M4\tablenotemark{11} &   10.97 &   10.31 &   10.10 & \nodata & \nodata & \nodata &       S  \\
  J2-157\dotfill &   4:20:52.7 &  17:46:41.5 & 151.92 &   M5.5\tablenotemark{9} &   11.61 &   11.04 &   10.78 & 0.7\tablenotemark{9} & -1.5\tablenotemark{9} & \nodata &  S  \\
  T~Tau\dotfill &   4:21:59.4 &  19:32:06.4 & 156.79 &         K0\tablenotemark{1} &    7.24 &    6.24 &    5.33 &     1.4 &    -1.2 &    5.73 &      BS  \\
  RX~J0422.1+1934\dotfill & 4:22:05.0 & 19:34:48.3 & 157.09 & M4.5\tablenotemark{11} & 10.10 &    9.05 &    8.69 & \nodata & \nodata & \nodata &       S  \\
  TAP~38\tablenotemark{6}\dotfill &   4:31:13.8 &  20:53:43.6 & 163.74 &   \nodata &    9.31 &    8.70 &    8.51 & \nodata & \nodata & \nodata &       S  \\
  RX~J0438.2+2023\dotfill & 4:38:13.0 & 20:22:47.1 & 164.15 & K2\tablenotemark{11} &   10.07 &    9.53 &    9.36 & \nodata & \nodata & \nodata &       S  \\
  RX~J0438.6+1546\dotfill & 4:38:39.1 & 15:46:13.7 & 177.46 & K1\tablenotemark{10} &    8.90 &    8.36 &    8.24 &     1.4 &    -1.9 & \nodata &       S  \\
  RX~J0438.4+1543\dotfill & 4:38:27.7 & 15:43:38.0 & 178.01 & K3\tablenotemark{11} &   10.92 &   10.31 &   10.14 &     0.7 &    -1.1 & \nodata &       S  \\
  TAP~28\tablenotemark{6}\dotfill &   4:19:17.3 &  17:14:59.4 & 181.60 &   \nodata &    9.96 &    9.36 &    9.18 & \nodata & \nodata & \nodata &       S  \\
  TAP~25\dotfill &   4:18:38.6 &  17:50:29.8 & 183.02 &        G0\tablenotemark{6} &    8.34 &    8.26 &    8.16 & \nodata & \nodata & \nodata &       S  \\
  RX~J0430.8+2113\dotfill & 4:30:49.2 & 21:14:10.6 & 184.36 & G8\tablenotemark{11} &    8.94 &    8.44 &    8.39 &     2.9 &    -2.8 & \nodata &      SM  \\
  HBC~376\dotfill &   4:18:51.7 &  17:23:16.6 & 185.09 &      K7\tablenotemark{8} &   10.03 &    9.42 &    9.27 &     0.9 &    -1.5 &    4.63 &       S  \\
  RX~J0423.7+1537\dotfill & 4:23:41.3 & 15:37:54.9 & 188.09 & K2\tablenotemark{10} &    9.41 &    8.93 &    8.81 &     0.8 &    -1.6 & \nodata &       S  \\
  HBC~372\dotfill &   4:18:21.5 &  16:58:47.0 & 199.62 &      K5\tablenotemark{8} &   11.18 &   10.60 &   10.46 &     1.1 &    -1.4 &    2.43 &       S  \\
  RX~J0444.4+1952\dotfill & 4:44:26.9 & 19:52:16.9 & 211.42 & M1\tablenotemark{11} &    9.56 &    8.88 &    8.71 &    -0.2 &     0.7 & \nodata &      SM  \\
  RX~J0431.3+2150\dotfill & 4:31:16.9 & 21:50:25.3 & 220.43 & K0\tablenotemark{11} &    9.24 &    8.83 &    8.73 &     0.3 &    -1.4 & \nodata &       S  \\
  RX~J0444.3+2017\dotfill & 4:44:23.6 & 20:17:17.5 & 223.75 & K1\tablenotemark{11} &   10.29 &    9.70 &    9.53 & \nodata & \nodata & \nodata &       S  \\
  RX~J0443.4+1546\dotfill & 4:43:26.0 & 15:46:03.9 & 224.76 & G7\tablenotemark{11} &   10.71 &   10.20 &   10.10 &     0.2 &    -1.5 & \nodata &       S  \\
  DQ~Tau\dotfill &   4:46:53.1 &  17:00:00.2 & 232.21 &        M0\tablenotemark{1} &    9.51 &    8.54 &    7.98 &     0.0 &    -0.6 &  -17.37 &      VS  \\
  Haro~6-37\dotfill &   4:46:59.0 &  17:02:38.2 & 232.75 &     K6\tablenotemark{1} &    9.24 &    7.99 &    7.31 &     0.0 &    -1.4 &    9.83 &      BS  \\
  Haro~6-37/c\dotfill &   4:46:59.0 &  17:02:38.2 & 232.75 &   K7\tablenotemark{1} & \nodata & \nodata & \nodata & \nodata & \nodata &    6.13 &      BS  \\
  DR~Tau\dotfill &   4:47:06.2 &  16:58:42.8 & 235.59 &
  K7\tablenotemark{1} &    8.85 &    7.80 &    6.87 &     0.5 &
  -1.7 & \nodata &      S 
\enddata
\tablenotetext{a}{From the 2MASS point source catalog unless otherwise
  cited.}
\tablenotetext{b}{Distance from \ceo\ intensity weighted mean position
  of L1551.}
\tablenotetext{c}{From \cite{duc05} unless otherwise cited. The units
  of measure for proper motion are arcsecond per century.}
\tablenotetext{d}{Radial velocities converted to the local standard of
  rest from \cite{her95}.}
\tablenotetext{e}{Note legend: ``B'' for members of known binary or
  multiple system; ``C'' for sources with anomalous near IR colors;
  ``E'' for embedded sources; ``M'' for sources with high proper
  motion relative to the mean of the sample;
  ``S'' for sources beyond $1.\!\!^\circ6$ from
  the L1551 center of mass; and ``V'' for sources with high radial
  velocity relative to the LSR velocity of the gas in L1551.}  
\tablerefs{1)~\cite{ken95}; 2)~\cite{ken98};
  3)~\cite{whi01}; 4)~\cite{mar99}; 5)~\cite{bri98};
  6)~\cite{fei87}; 7)~\cite{whi99}; 8)~\cite{her95};
  9)~\cite{gom92}; 10)~\cite{wic00}; 11)~\cite{wic96};
  12)~\cite{bas95}; 13)~\cite{sar98}}
\label{PMStable}
\end{deluxetable}

\renewcommand{\arraystretch}{1}

\newpage

\renewcommand{\arraystretch}{1.4}

\begin{center}

\begin{deluxetable}{lcccccccc}
\tablecolumns{9}
\tablewidth{0pt}
\tablecaption{Properties of Selected Pre-main Sequence
  Stars\label{PMSprops}}
\tablehead{\colhead{Name} &
\colhead{$T_{\rm eff}$ (K)} &
\colhead{$L_{\rm bol} (L_\odot)$} &
\colhead{Type} &
\colhead{Class} &
\colhead{$A_V$} &
\colhead{$J_{\rm c}$} &
\colhead{age (Myr)} &
\colhead{$M$ (\msun)} }
\startdata
L1551~IRS~5 (A+B)\dotfill &  5000 & 21.6\tablenotemark{1} & proto & I &    9.70 &   10.97     & $ 0.1^{+ 0.9}_{- 0.1}$ & \nodata  \\
HH30~IRS\dotfill &     3850 & 0.004 &  C &   I/II                     & \nodata &   15.18     & $ 0.1^{+ 0.9}_{- 0.1}$ & \nodata  \\
LkH$\alpha$~358\dotfill & 3058 & 0.24 & C &    II                     &    8.82 &   10.30     & $ 0.1^{+ 0.9}_{- 0.1}$ & \nodata  \\
 L1551~NE (A+B)\dotfill & \nodata & 4.70\tablenotemark{1} & proto & 0 & \nodata &   16.61     & $ 0.1^{+ 0.9}_{- 0.1}$ & \nodata  \\
   HL~Tau\dotfill &    4060 &  0.26 & C &      II                     & \nodata &   10.62     & $ 0.1^{+ 0.9}_{- 0.1}$ & \nodata  \\
  XZ~Tau~A\dotfill &   3415 &  1.24 & C &      II                     &    3.07 &    8.52     & $ 0.1^{+ 0.9}_{- 0.1}$ & \nodata  \\
  XZ~Tau~B\dotfill &   3633 &  0.05\tablenotemark{2} & C & \nodata    &    1.39 & \nodata     & $ 0.1^{+ 0.9}_{- 0.1}$ & \nodata  \\
    MHO~4\dotfill &    2990 &  0.08 & W & \nodata                     &    0.69\tablenotemark{3} &   11.46     & $ 1.0^{+ 0.8}_{- 0.2}$ &    0.11  \\
    MHO~9\dotfill &    3270 &  0.18 & W & \nodata                     &    1.73\tablenotemark{3} &   10.72     & $ 4.0^{+ 0.5}_{- 2.4}$ &    0.25  \\
    MHO~5\dotfill &    2990 &  0.12 & C &      II                     &    0.11\tablenotemark{3} &   11.04     & $ 1.3^{+ 0.3}_{- 0.3}$ &    0.17  \\
V~710~Tau~B\dotfill &  3560 &  0.63\tablenotemark{2} & W &      II    &    1.82\tablenotemark{2} &    8.47     & $ 1.3^{+ 0.8}_{- 0.3}$ &    0.62  \\
V~710~Tau~A\dotfill &  3705 &  0.51\tablenotemark{2} & C &      II    &    1.80\tablenotemark{2} &    8.77     & $ 3.2^{+ 1.8}_{- 0.9}$ &    0.80  \\
 V~826~Tau\dotfill &   4060 &  1.30 & W &     III                     &    0.69\tablenotemark{4} &    8.88     & $ 2.8^{+ 2.2}_{- 1.4}$ &    1.20  \\
 V~827~Tau\dotfill &   4060 &  1.33 & W &     III                     &    1.14\tablenotemark{4} &    8.85     & $ 2.5^{+ 2.5}_{- 0.7}$ &    1.20  \\
L1551-51\dotfill &     4060 &  0.61 & W &     III                     &    0.00 &    9.70     & $ 6.4^{+ 2.7}_{- 2.8}$ &    1.10  \\
 RX~J0432.6+1809\dotfill &  3125 & 0.10 & W & \nodata                 &    0.00 &   11.31     & $ 2.2^{+ 0.6}_{- 0.8}$ &    0.20  \\
  UX~Tau~A\dotfill &   4350 &  1.93 & C &      II                     &    0.26\tablenotemark{2} &    8.55     & $ 1.8^{+ 1.8}_{- 0.8}$ &    1.40  \\
  UX~Tau~B\dotfill &   3560 &  0.19 & W &     III                     &    0.26\tablenotemark{2} &    9.80     & $ 7.1^{+ 1.9}_{- 3.1}$ &    0.57  \\
  UX~Tau~C\dotfill &   3125 &  0.10\tablenotemark{2} & W & \nodata    &    0.57\tablenotemark{2} & \nodata     & $ 1.3^{+ 2.3}_{- 0.3}$ &    0.13  \\
L1551-55\dotfill &     4060 &  0.52 & W &     III                     &    0.69\tablenotemark{1} &    9.97     & $ 8.9^{+ 3.7}_{- 3.9}$ &    1.05  \\
    MHO~6\dotfill &    3125 &  0.08 & W & \nodata                     &    0.54\tablenotemark{3} &   11.56     & $ 2.8^{+ 1.2}_{- 0.6}$ &    0.17  \\
    MHO~7\dotfill &    3125 &  0.11 & W & \nodata                     &    0.23\tablenotemark{3} &   11.12     & $ 1.6^{+ 0.9}_{- 0.5}$ &    0.20  \\
   DM~Tau\dotfill &    3705 &  0.33 & C &      II                     &    0.00 &   10.44     & $ 6.4^{+ 4.9}_{- 2.4}$ &    0.70  \\
  HN~Tau~A\dotfill &   4350 &  0.49 & C &      II                     &    2.79 &   10.03     & $20.0^{+11.7}_{- 7.3}$ &    1.05  \\
  HN~Tau~B\dotfill &   3270 &  0.03\tablenotemark{2} & C & \nodata    &    0.91\tablenotemark{2} &   12.36     & $11.4^{+ 8.8}_{- 2.5}$ &    0.17  \\
 RX~J0432.8+1735\dotfill &   3560 &  0.38 & W & \nodata               &    0.00 &   10.00     & $ 2.8^{+ 0.3}_{- 1.3}$ &    0.62  \\
 GG~Tau~Aa\dotfill &   4060 &  1.52 & C &      II                     &    1.90 &    8.70     & $ 1.6^{+ 1.6}_{- 0.6}$ &    1.20  \\
 GG~Tau~Ab\dotfill &   3778 &  0.82 & C &      II                     &    3.01 &    9.27     & $ 1.8^{+ 1.4}_{- 0.6}$ &    0.90  \\
 GG~Tau~Ba\dotfill &   3125 &  0.11 & C &      II                     &    1.12 &   11.16     & $ 1.0^{+ 1.0}_{- 0.1}$ &    0.13  \\
 GG~Tau~Bb\dotfill &   2880 &  0.02 & C &      II                     &    0.63 &   12.98     & $ 2.5^{+ 0.9}_{- 0.5}$ &    0.05  \\
  HBC~407\dotfill &    5520 &  0.71 & W &     III                     &    1.95 &   10.03     &                    MS? &    1.00  \\
 J2-2041\dotfill &     3343 &  0.25 & W &     III                     &    0.90\tablenotemark{5} &   10.28     & $ 1.8^{+ 2.2}_{- 0.1}$ &    0.30  \\
 RX~J0432.7+1853\dotfill &  5080 &  1.23 & W & \nodata                &    0.00 &    9.24     & $22.4^{+22.6}_{- 6.6}$ &    1.15  \\
   TAP~51\dotfill &    6200 &  2.08 & W & \nodata                     &    0.91 &    9.06     &                    MS? &    1.20  \\
  HBC~412\dotfill &    3560 &  0.37 & W &     III                     &    0.00 &   10.03     & $ 2.5^{+ 2.0}_{- 0.9}$ &    0.57  \\
  HBC~388\dotfill &    5080 &  1.87 & W & \nodata                     &    0.00 &    8.78     & $12.6^{+ 9.8}_{- 1.5}$ &    1.40  \\
  HBC~392\dotfill &    4350 &  0.74 & W &     III &    2.46 &    9.59     & $12.6^{+ 1.6}_{- 6.9}$ &    1.20 
\enddata
\tablerefs{1)~\cite{ken95}; 2)~\cite{whi01}; 3)~\cite{bri98};
  4)~\cite{luh00}; 5)~\cite{gom92} }
\end{deluxetable}

\end{center}

\renewcommand{\arraystretch}{1}

\end{document}